\documentclass[pre,showpacs,twocolumn,floatfix]{revtex4-1}
\usepackage{multirow}
\usepackage{amsmath}
\usepackage{graphicx}
\usepackage{comment}
\usepackage{float}
\usepackage{amssymb}
\usepackage{times}
\usepackage[colorlinks,hyperindex]{hyperref}
\usepackage{subfigure}
\hypersetup
	{
		colorlinks,%
		citecolor=blue,%
		linkcolor=blue,%
		urlcolor=black,%
	}

\begin{document}
\graphicspath{{graphics/}}
\title{Kibble-Zurek Mechanism and Finite-Time Scaling}

\author{Yingyi Huang}

\author{Shuai Yin}

\author{Baoquan Feng}

\author{Fan Zhong}  \thanks{Corresponding author. E-mail: stszf@mail.sysu.edu.cn}
\affiliation{State Key Laboratory of Optoelectronic Materials and
Technologies, School of Physics and Engineering, Sun Yat-sen
University, Guangzhou 510275, People's Republic of China}

\date{\today}

\begin{abstract}
The Kibble-Zurek (KZ) mechanism has been applied to a variety of systems ranging from low temperature Bose-Einstein condensations to grand unification scales in particle physics and cosmology and from classical phase transitions to quantum phase transitions. Here we show that finite-time scaling (FTS) provides a detailed improved understanding of the mechanism. In particular, the finite time scale, which is introduced by the external driving (or quenching) and results in FTS, is the origin of the division of the adiabatic regimes from the impulse regime in the KZ mechanism. The origin of the KZ scaling for the defect density, generated during the driving through a critical point, is not that the correlation length ceases growing in the nonadiabatic impulse regime, but rather, is that it is taken over by the effective finite length scale corresponding to the finite time scale. We also show that FTS accounts well for and improves the scaling ansatz proposed recently by Liu, Polkovnikov, and Sandvik [Phys. Rev. B {\bf 89}, 054307 (2014)]. Further, we show that their universal power-law scaling form applies only to some observables in cooling but not to heating. Even in cooling, it is invalid either when an appropriate external field is present. However, this finite-time-finite-size scaling calls for caution in application of FTS. Detailed scaling behaviors of the FTS and finite-size scaling, along with their crossover, are explicitly demonstrated, with the dynamic critical exponent $z$ being estimated for two- and three-dimensional Ising models under the usual Metropolis dynamics. These values of $z$ are found to give rise to better data collapses than the extant values do in most cases but take on different values in heating and cooling in both two- and three-dimensional spaces.
\end{abstract}

%---------------------------------------------------------------
%---------------------------------------------------------------
\maketitle

\section{Introduction}
\label{sec:intro}

The Kibble-Zurek (KZ) mechanism, the mechanism for defect formations when a system is driven (or quenched in the context) through a continuous phase transition into an ordered state, has received a lot of attentions for many years. This mechanism was first proposed by Kibble in cosmology. It predicts that topological defects can be formed with cosmological significance as the universe expands and cools through a continuous phase transition in a spontaneously broken gauge theory~\cite{kibble1}. Later Zurek applied the critical scalings of the relaxation time and correlation length in the transition to compute the nonequilibrium scaling of defect density in condensed-matter physics~\cite{zurek1}. This predication agrees with many numerical simulations and experiments~\cite{KZtest} except in superfluid $^4$He~\cite{He}. The KZ mechanism has also been applied successfully to quantum phase transitions at least through a single quantum critical point either linearly or nonlinearly~\cite{zurekq,deng}. See Ref.~[\onlinecite{Dziarmaga}] for reviews.

Recently, the KZ Mechanism has been applied back to classical critical phenomena to determine nonequilibirum scaling at the critical point by Liu, Polkovnikov, and Sandvik (LPS)~\cite{Liu}. They proposed two scaling functions governing dynamic scalings in the so-called adiabatic and diabatic regimes. Both functions cross over into the same universal power law in a universal scaling regime. These dynamic scalings were demonstrated beautifully by a dynamic finite-size scaling (FSS) using different types of stochastic dynamics on Ising models. A new estimate of the dynamic critical exponent $z$ for the two-dimensional (2D) Ising model was also obtained.

On the other hand, we have applied a linearly varying external field~\cite{Zhongjp,Zhongb05,Zhong06} and temperature~\cite{Zhonge1} to study scaling in classical first-order phase transitions~\cite{Zhongjp,Zhonge1}, classical continuous phase transitions~\cite{Zhongb05,Zhong06,Gong10}, and recently quantum phase transitions~\cite{Yin12,Yin13} following similar studies using internal frictions~\cite{Zhang}. Renormalization-group theories both for a field driving and a temperature driving have also been developed to derive the scaling with the rate of the sweeping~\cite{Zhong06}. A theory of finite-time scaling (FTS) has also been proposed and confirmed numerically~\cite{Gong10} and crossover from FTS to FSS has been considered theoretically but not yet been tested numerically~\cite{Zhong11}. However, these FTS forms are different from the scaling proposed by LPS in Ref.~[\onlinecite{Liu}]. Accordingly, a detailed comparison is desirable.

Here, we shall first show that FTS provides a detailed improved understanding of the KZ Mechanism. In particular, the finite time scale, which is introduced by the external driving (or quenching) and results in the concept of FTS, is the origin of the division of the adiabatic regimes from the impulse regime in the KZ Mechanism. The origin of the KZ scaling for the defect density generated during the driving is not that the correlation length ceases growing in the nonadiabatic impulse regime, but rather, is that it is taken over by the effective finite length scale corresponding to the finite time scale. In fact, the impulse regime is just the critical region in which scaling originates and is just the FTS regime. So, the KZ scaling just marks the crossover of the FTS regime to the equilibrium regimes, which is the adiabatic regimes of the KZ mechanism.

We shall then show that FTS can not only well account for the scaling proposed by LPS, which incorporates the KZ mechanism with FSS, but also improves it by making up its missing subleading contributions. Thus, first, the LPS adiabatic regime is just the usual FSS regime in which the lattice size is the controlling length scale. As a result, it is also universal. Second, the LPS universal scaling regime is just the FTS regime in which the externally imposed time scale is the controlling time scale. In this FTS regime, the dynamics is nonadiabatic or diabatic and the system falls out of equilibrium similar to the LPS diabatic regime. Third, the LPS scaling function for the diabatic regime is a different representation of the FTS form. Further, we find that the LPS universal scaling is applicable only to some observables in cooling but not to heating as there exist restrictions to fluctuations that render cooling and heating different. Even in cooling, this scaling is invalid either when the system is subjected to a small external field which, if large enough, can also remove these restrictions. However, for not sufficiently large applied fields, the LPS scaling may appear in an intermediate regime between the FSS and the standard FTS regimes. For differentiation, we thus call it as a finite-time-finite-size scaling regime. In addition, as the LPS universal scaling regime is essentially the impulse regime, applying the finite-size KZ mechanism to explain the scaling at the critical point that lies deep in the impulse regime is essentially inconsistent, since the impulse regime is assumed in the KZ mechanism as evolutionless.

We shall finally study in detail FTS and FSS and their crossover for several observables in both FTS and FSS forms and unambiguously confirm them numerically. The influence of the LPS finite-time-finite-size scaling on the application of FTS will be demonstrated. The dynamic critical exponent $z$ will be estimated for the 2D and 3D Ising models with the usual Metropolis dynamics using FTS and FSS and their crossover. These estimated values give rise to better data collapses than the extant values do in most cases but take on different values in heating and cooling in both 2D and 3D. Corrections to scaling are found to be prominent in heating for the extant values.

In the following, we first review the KZ mechanism and the scaling proposed by LPS in Sec.~\ref{kz} and FTS in Sec.~\ref{fts}, which allow us to reveal their relationship in Sec.~\ref{relation}. We then summarize the characteristics of the FSS and FTS graphs for several observables in Sec.~\ref{chara} and test them with Monte Carlo simulations from 2D and 3D classical Ising models in Sec.~\ref{results}. Various estimates of the dynamic critical exponents are presented there. Finally, conclusions are given in Sec.~\ref{sum}.

\section{\label{kz}KZ mechanism and LPS scaling}
The KZ mechanism facilitates finding the density of defects formed when a system is driven by varying a parameter through its critical value. For comparison with LPS~\cite{Liu}, consider cooling a system from an initial temperature $T_i$ through the critical temperature $T_c$ with a constant rate $R$ ($v$ in LPS). In the vicinity of $T_c$, the system is characterized by a large correlation length $\xi$ and a correlation or relaxation time $t_{\rm eq}$, both of which diverge at the critical point as
\begin{eqnarray}
\xi&\sim& \tau^{-\nu},\label{xi}\\
t_{\rm eq}&\sim& \xi^{z}\sim\tau^{-\nu z},\label{zeta}
\end{eqnarray}
respectively, where the reduced temperature $\tau=T-T_c$ and $\nu$ is the correlation-length critical exponent~\cite{Cardy}. So, no matter how small the cooling rate $R$ is, there exists a frozen instant $\hat{t}$ at which the system cannot follow the cooling adiabatically and falls out of equilibrium, because the relaxation time of the system is longer than the time when the temperature is changed. To find $\hat{t}$, note that
\begin{equation}
\tau=Rt,\label{trt}
\end{equation}
because the time origin can always be chosen in a way such that at $t=0$, $T=T_c$, independent of $T_i$ for a linear cooling. So, at $\hat{t}$, ``the remaining time until the transition''~\cite{zurek1,KZtest}, the corresponding frozen reduced temperature $\hat{\tau}$ satisfies $\hat{\tau}/R=t_{\rm eq}\sim\hat{\tau}^{-\nu z}$, i.e.,
\begin{equation}
\hat{\tau}\sim R^{1/(1+\nu z)}.\label{hattau}
\end{equation}
This leads back to
\begin{equation}
\hat{t}\sim R^{-\nu z/(1+\nu z)}.\label{hatt}
\end{equation}
The KZ mechanism thus assumes that $\xi(t)$ will not grow further once $\hat{t}$ is reached, which may result from the assumed evolutionless of the state of the system in the quantum context~\cite{zurekq,Dziarmaga}. So regions of about $\hat{\xi}$, with
\begin{equation}
\hat{\xi}\equiv\xi(\hat{t})\sim R^{-\nu/(1+\nu z)},\label{hatxi}
\end{equation}
apart can mimic the causality-independent region in cosmology and topological defects can be found if the homotopy group of the order parameter is nontrivial~\cite{kibble1,zurek1}. Accordingly, the generated defect density $n$ is proportional to $\hat{\xi}^{-d}$, or,
\begin{equation}
n\sim R^{d\nu/(1+\nu z)}\label{dd}
\end{equation}
which is the KZ scaling. The system is frozen until $-\hat{t}$ at which (quasi-)adiabatic evolution reassumes, but the correlation length then decreases and the defects survive.

One sees therefore that the KZ mechanism divides the evolution of the system during the cooling into three regimes separated by $\pm\hat{t}$. Two adiabatic regimes at both ends are separated by an impulse regime that embraces the critical point and is assumed to be dark as the system is assumed to cease evolving there. The KZ scaling, (\ref{dd}), is just a characteristic of the borders of the impulse regime. In addition, the KZ division of different regimes has methodological implications. In the adiabatic regimes, perturbation expansions work, while in the impulse regime, non-perturbative methods have to be employed.

In order to combine the KZ mechanism with the standard FSS, note first that Eq.~(\ref{hattau}) can be reversed and defines a general (not specific to $\hat{\tau}$) KZ rate $R_{\rm KZ}$ as
\begin{equation}
R_{\rm KZ}\sim\tau^{1+\nu z}.\label{rkz}
\end{equation}
For $R$ larger than $R_{\rm KZ}$, the system will fall out of equilibrium at $T$. At $T=T_c$ or $\tau=0$, $R_{\rm KZ}=0$. One sees therefore that the system cannot remain adiabatic all the way down to $T_c$ for any finite rate $R$ as expected. However, this is possible for a finite-size system. Reversing Eq.~(\ref{hatxi}) and replacing $\hat{\xi}$ with the largest length scale of a finite system, its length $L$, one defines a size-dependent KZ rate $R_{\rm KZ}(L)$ as~\cite{Liu}
\begin{equation}
R_{\rm KZ}(L)\sim L^{-(z+1/\nu)}.\label{rkzl}
\end{equation}
Note that this is the rate at which the correlation length $\hat{\xi}$ at the frozen reduced temperature $\hat{\tau}$ is of the order of the system size $L$. Consequently, for $R$ smaller than $R_{\rm KZ}(L)$, the system remains adiabatic all the way down to $T_c$ and perturbative treatments are possible, as the correlation length has already larger than the system size. On the other hand, for $R$ larger than $R_{\rm KZ}(L)$, the system will be frozen before it reaches $\hat{\tau}$ and so adiabaticity breaks down.

LPS~\cite{Liu} considered the cooling of Ising models to their critical points and studied the scaling behavior of the averaged squared magnetization $\langle m^2 \rangle$, where the angle brackets stand for ensemble averages over different random-number sequences. Using the techniques of the standard FSS~\cite{Barber1983}, they assumed that the argument $L/\hat{\xi}$ should enter the scaling for a linear driving in addition to the equilibrium argument $L/\xi$ and wrote a scaling ansatz for $\langle m^2 \rangle$ as~\cite{deng}
\begin{equation}
\label{m2scaling}
\langle m^2 \rangle = L^{-2 \beta/\nu} F_1 \big( \tau L^{1/\nu}, R L^{z+1/\nu} \big),
\end{equation}
where $F_1$ is a scaling function. For simplification, they measured $\langle m^2 \rangle$ at $T=T_c$. Consequently, the first argument of $F_1$ vanishes and so
\begin{equation}
\label{m2scaling1}
\langle m^2 \rangle = L^{-2 \beta/\nu} F_{10} \big( R L^{z +1/\nu} \big),
\end{equation}
where $F_{10}(Y)\equiv F_1(0,Y)$, a rule which will be followed throughout.

More detailed scaling behavior can then be found from the scaling ansatz~(\ref{m2scaling1}) along with other plausible arguments. For small $R$, the standard FSS,
\begin{equation}
   \label{fsscale}
\langle m^2\rangle \sim L^{-2\beta/\nu},
\end{equation}
must recover, while for sufficiently large $R$,
\begin{equation}
\langle m^2 \rangle \sim L^{-d}\label{md}
\end{equation}
and depends on $T_i$ since the initial state hardly evolves for large $R$~\cite{Liu}. This then prompts LPS to demand that $F_{10}(Y)$ must reduce to a pure power law of its argument with an exponent,
\begin{equation}
   \label{power}
   x = \dfrac{ d - 2 \beta/\nu  }{ z + 1/\nu },
\end{equation}
such that
\begin{equation}
   \label{m2vx}
    \langle m^2 \rangle \sim  L^{-d}R^{-x}.
\end{equation}
in an intermediate universal scaling regime~\cite{Liu}, where $\beta$ is the critical exponent associated with the averaged magnetization $\langle m\rangle$. As the scaling~(\ref{m2vx}) is obviously not consistent with the large rate limit for fixed $L$, (\ref{md}), LPS proposed to replace $F_1$ by two scaling functions, introduced further a size-independent upper limit $R_a\sim a^{-(z+1/\nu)}$ beyond which the power-law behavior (\ref{m2vx}) breaks down, and wrote
\begin{equation}
 \label{2scaling_functions}
  \langle m^2 \rangle = \left \{
\begin{array}{l l}
L^{-2\beta/\nu} a^{-d+2\beta/\nu} \hspace{1pt} F_{20} \big( R L^{z+1/\nu} \big),  &  R < R_{{KZ}}(L)\\
& \\
L^{-d} \hspace{1pt} F_{30} \big(a^{-(z+1/\nu)}R^{-1} \big),  & R > R_a ,
\end{array}
\right.
\end{equation}
where $a$ is a short-range length scale, which is of the order of one lattice spacing, and is mainly for dimension reasons. Both scaling functions $F_{20}$ and $F_{30}$ cross over to the power law~(\ref{m2vx}) in the intermediate universal scaling regime $ R_{{KZ}}(L) < R < R_a$. LPS' numerical results show that $F_{20}$ characterizes well the scaling in the quasi-adiabatic and universal scaling regimes, whereas $F_{30}$ describes only the universal scaling regime but not the large-rate diabatic regime~\cite{Liu}. We shall show in the following that the power-law behavior~(\ref{m2vx}) is only true for some observables in cooling and under no or a small applied external field. We shall also improve it by completing its full scaling form.

\section{\label{fts}FTS}
FTS is a temporal analogue of the FSS. FSS is a method to circumvent the embarrassment that the size $L$ of a system can even fall behind the correlation length $\xi$ inevitably in the neighborhood of a critical point. But the same nuisance happens in the time domain because of Eq.~(\ref{zeta}) and is the notorious critical slowing down. FTS adopts the tactics of FSS by devising a variable time scale that is readily controllable like $L$~\cite{Gong10,Zhong11}.

To see this, consider the dynamic scale transformation of the susceptibility defined as
\begin{equation}
\chi=L^d\left(\langle m^2\rangle-\langle m\rangle^2\right),\label{chi}
\end{equation}
where we have neglected a temperature factor as we shall always work at the critical point. Near a critical point and in the thermodynamic limit $L\rightarrow\infty$~\cite{Hohenberg,Justin,Cardy,Gong10,Zhong11},
\begin{equation}
\chi(\tau,t) =b^{\gamma/\nu}\chi\left(\tau b^{1/\nu},tb^{-z}\right),\label{chit}
\end{equation}
where $b$ is a scale factor and $\gamma$ the critical exponent for $\chi$. For a \emph{constant} $\tau$, Eq.~(\ref{chit}) is well established within the renormalization-group theory for critical dynamics~\cite{Hohenberg,Justin,Cardy}. Choosing $b$ on such a scale that $\tau b^{1/\nu}$ is of order unity, one finds from the second argument of Eq.~(\ref{chit}) the correlation time $t_{\rm eq}$ in Eq.~(\ref{zeta}). $t_{\rm eq}$ is finite for a finite $\tau$ and one could then estimate critical properties by finding $\chi$ for sufficiently long time. However, a more useful strategy is to change $\tau$ with time. If Eq.~(\ref{chit}) is valid for a \emph{time-dependent} $\tau$ in the linear protocol~(\ref{trt}), the similar procedure gives rise to another time scale
\begin{equation}
t_R\sim R^{-z/r},\label{tr}
\end{equation}
which is constant for a constant $R$ in contrast to the varying $t_{\rm eq}$ with $\tau$, where $r$ is the renormalization-group eigenvalue of $R$ defined as $R'=Rb^r$ upon rescaling. Equation~(\ref{tr}) can be derived by replacing the first argument of $\chi$ on the right hand side of Eq.~(\ref{chit}) with $R'$ and choosing $b$ on the scale of $R^{-1/r}$. The reason of the replacement is that $\tau=Rt$ and so only two out of the trio are independent. From the rescaling of Eq.~(\ref{trt}), viz. $\tau'=R't'$ (this can be considered to be a definition of $R'$), with $\tau'=\tau b^{1/\nu}$ and $t'=tb^{-z}$ from Eq.~(\ref{chit}), one finds
\begin{equation}
r=z+1/\nu,\label{r}
\end{equation}
not to be confused with the nonlinear exponent used by LPS~\cite{Liu}. In fact, for nonlinear protocols of $\tau=Rt^n$ with a constant $n$, the above method yields simply~\cite{Gong10}
\begin{equation}
r_n=nz+1/\nu\label{rn}
\end{equation}
in agreement with LPS~\cite{Liu}.
Therefore, one sees that varying linearly $\tau$ does give rise to a new constant time scale $t_R$. It is readily controllable simply by using, say, different cooling rates.

A renormalization-group theory using the technique of composite operator insertions~\cite{Justin,Amit} has been developed to justify the validity of Eq.~(\ref{chit}) for a time-dependent $\tau$~\cite{Zhong06}. Within the framework of field-theoretical renormalization-group theory~\cite{Justin,Amit}, no genuine new exponents have to be introduced in the nonequilibrium driving process across the critical point, as the linear driving introduces no intrinsic new divergences to the original theory.

With the controllable finite time scale, FTS then follows. Replacing the last argument on the right hand side of Eq.~(\ref{chit}) by $Rb^r$, for example, and choosing $b$ on the scale of $R^{-1/r}$, one finds an FTS form
\begin{equation}
\chi=R^{-\gamma/r\nu}\tilde{f}\left(\tau R^{-1/r\nu}\right)\label{chifts}
\end{equation}
with a scaling function $\tilde{f}$. The similarity to FSS can be clearly seen here. The FTS regime shows up for $\tau R^{-1/r\nu}\ll1$, which is just $t_R\ll t_{\rm eq}$ as expected. On the other hand, equilibrium behavior $\chi\sim\tau^{-\gamma}$ follows in the other equilibrium regime. This is just the logic of FSS~\cite{Barber1983}.

If the size of the system is finite, Eq.~(\ref{chit}) can be extended to
\begin{equation}
\chi(\tau,t,L) =b^{\gamma/\nu}\chi\left(\tau b^{1/\nu},tb^{-z},L^{-1}b\right)\label{chitl}
\end{equation}
as is usually done in FSS. Then choosing suitable scale factors leads to an FTS form~\cite{Zhong11}
\begin{equation}
\chi=R^{-\gamma/r\nu}f\left(\tau R^{-1/r\nu},L^{-1}R^{-1/r}\right)\label{chiftsl}
\end{equation}
or an FSS form
\begin{equation}
\chi=L^{\gamma/\nu}f_1\left(\tau L^{1/\nu},R L^{r}\right),\label{chifssr}
\end{equation}
where $f$ and $f_1$ are scaling functions and are analytic when their arguments are small. Note that both scaling forms can describe both the FTS regime and the FSS regime similar to $F_{20}$ in Eq.~(\ref{2scaling_functions}), which can describe both the adiabatic and the universal scaling regimes. In fact, one can check that
\begin{equation}
f(X,Y)=Y^{-\gamma/\nu}f_1\left(XY^{-1/\nu},Y^{-r}\right).\label{f12}
\end{equation}
Besides the usual conditions that in the FTS regime, $t_R<t_{\rm eq}$ and in the FSS regime, $L<\xi$, Eqs.~(\ref{chiftsl}) and (\ref{chifssr}) also require roughly that in the former regime, $R > R_{\rm KZ}(L)$ and in the latter, $R < R_{\rm KZ}(L)$ by using Eqs.~(\ref{rkzl}) and (\ref{r}), i.e., high rates for the FTS and lower rates for the FSS. Using Eq.~(\ref{hatxi}), which, in fact, defines an effective length scale associated with the driving, we can express these conditions in terms of length scales. Thus, from Eq.~(\ref{chiftsl}), the FTS regime has $\xi>\hat{\xi}$ and $L>\hat{\xi}$, which mean reasonably that the effective length scale is the shortest among $\hat{\xi}$, $L$, and $\xi$; while in the FSS, $L$ is the shortest.

\section{\label{relation}Relationship between FTS and KZ mechanism and LPS scaling}

\subsection{\label{fk}FTS vs KZ mechanism}
In this section, we summarize several relations between FTS and the KZ mechanism from the foregoing review.

From FTS, (a) the reason that the frozen instant $\hat{t}$, at which the remaining time until the transition equals the correlation time $t_{\rm eq}$~\cite{zurek1,KZtest}, is the division of the adiabatic and the impulse regime becomes clear. Note that Eqs.~(\ref{tr}) and (\ref{r}) is just the frozen instant $\hat{t}$ in Eq.~(\ref{hatt}). So, the instant at $\hat{t}$ just equals the externally imposed time scale $t_R$ and thus at this instant $t_R=t_{\rm eq}$. As a result, within the impulse regime, $t_R$ remains constant but $t_{\rm eq}$ increases and thus the external time scale is shorter than the relaxation time and the system falls out of equilibrium. Therefore, by identifying the finite time scale, the KZ division of different regimes is manifest.

Accordingly, (b) it is clear that the impulse regime is just the regime of FTS and thus the system does not cease evolving. A similar conclusion has been reached in quantum phase transitions~\cite{Yin12}. The situation here is similar to the FSS regime in which the size of the system $L$ gets shorter than the correlation length $\xi$. In this case, although the intrinsic correlation length may be still growing according to Eq.~(\ref{xi}), the system can only feel the correlation length to the extent of its size $L$.

Therefore, (c) although the defect density is still determined by $\hat{\xi}$, this is not because the intrinsic correlation length $\xi$ does not grow or does not subject to substantially grow, but because it is taken over by the driving: The constant driving time scale $t_R$ corresponds to a constant length scale $\hat{\xi}$, which is smaller than $\xi$.

In fact, (d) in comparison with the KZ mechanism, it is the evolution of system within the FTS regime that leads to the scaling, because this is the critical regime that embraces the critical point that in turn is the origin of the scaling.

So, (e) the KZ scaling only marks the crossover from the FTS regime to the equilibrium regime. Indeed, from Eq.~(\ref{chifts}), the crossover occurs at $\tau R^{-1/r\nu}\sim1$, which just results in Eq.~(\ref{hattau}).

\subsection{\label{fl}FTS vs LPS' scaling}
We study the relation between FTS and the LPS scaling~\cite{Liu} in this section.

First, similar to the conclusion (d) in Sec.~\ref{fk}, $R_{\rm KZ}(L)$ in Eq.~(\ref{rkzl}) just marks the border or crossover between the adiabatic FSS regime and the impulse FTS regimes of a finite-size system. Indeed, as mentioned, from Eqs.~(\ref{chiftsl}) and (\ref{chifssr}), the crossover takes place in a region at which the last arguments on the right hand sides of Eqs.~(\ref{chiftsl}) and (\ref{chifssr}) are of order unity, viz., $R\sim R_{\rm KZ}(L)$. From Sec.~\ref{fts}, we know that in the FSS regime, $R\ll R_{\rm KZ}(L)$ and $L$ is shorter than the driving length scale $\hat{\xi}$. So, only in this regime can the system remain adiabatic all the way down to $T_c$. In the FTS regime, $R> R_{\rm KZ}(L)$ or $\hat{\xi}<L$ and the system falls out of equilibrium inevitably. Therefore, although Eq.~(\ref{m2scaling}), which is just Eq.~(\ref{chifssr}) for another observable, may be regarded as bridging between the adiabatic and impulse regimes, its inherent KZ mechanism nature implies an inconsistency in applying the KZ scaling at the border to $T_c$ deep in the impulse regime.

Next, LPS' peculiar scaling form~\cite{Liu} is just the leading contribution of FTS. At $\tau=0$, Eqs.~(\ref{chifssr}) and (\ref{chiftsl}) becomes
\begin{equation}
 \label{ftsrl}
  \chi = \left \{
\begin{array}{l l}
L^{\gamma/\nu} f_{10} \left(R L^{r} \right),  &  R < R_{\rm KZ}(L)\\
& \\
R^{-\gamma/r\nu} f_{0} \left(L^{-1}R^{-1/r}\right),  & R > R_{\rm KZ}(L),
\end{array}
\right.
\end{equation}
respectively. Note that
\begin{equation}
d\nu=\gamma+2\beta\label{dn}
\end{equation}
from the scaling laws~\cite{Cardy}
\begin{eqnarray}
2-d\nu&=&\alpha,\nonumber\\
\alpha+2\beta+\gamma&=&2,\label{sc}
\end{eqnarray}
where $\alpha$ is the specific-heat critical exponent. So,
\begin{equation}
x=\gamma/r\nu.\label{xgamma}
\end{equation}
Combining Eq.~(\ref{dn}) with Eqs.~(\ref{chi}) (note that $\langle m \rangle=0$ in principle but see below for details) and (\ref{r}), one can convince oneself that the first line of Eq.~(\ref{ftsrl}) is just the corresponding one in Eq.~(\ref{2scaling_functions}) and describes the FSS regime, while the second line is the complete form of Eq.~(\ref{m2vx}) and describes the FTS regime.

Therefore, we have obtained the LPS scaling from FTS and shown that it is just a form of FTS, because just in this regime in which $\chi$ exhibits the standard FTS form, $\langle m^2 \rangle$ shows the LPS scaling. This will become clearer in the following. We see that the $L^d$ factor in the definition of $\chi$ automatically gives rise to the size factor of the scaling~(\ref{m2vx}) in FTS, though no new factor is produced in FSS. Moreover, we have improved the LPS scaling by including the subleading contribution from $L^{-1}$ in the FTS regime in the second line of Eq.~(\ref{ftsrl}). As a consequence of this subleading contribution, the first line of Eq.~(\ref{ftsrl}) is now not a pure power law in LPS' universal scaling regime as exhibited in Eqs.~(\ref{m2scaling1}) to (\ref{m2vx}), but is given by Eq.~(\ref{f12}).

We have utilized $\chi$ to derive and complete the LPS scaling. We can of course do the same thing from the squared magnetization itself. We shall see that this provides both a physical explanation for and the limitations of the LPS scaling.

To this end, we first write the averaged magnetization and its squared themselves in the FTS forms. They are~\cite{Zhong11}
\begin{eqnarray}
\langle m\rangle &=& R^{\beta/r\nu}{\cal F}\left(\tau R^{-1/r\nu},L^{-1}R^{-1/r}\right),\label{mftsr}\\
\langle m^2\rangle &=& R^{2\beta/r\nu}F\left(\tau R^{-1/r\nu},L^{-1}R^{-1/r}\right),\label{m2ftsr}
\end{eqnarray}
respectively. One can readily confirm that the scaling functions ${\cal F}$ and $F$ are related to their FSS counterparts of
\begin{equation}
\label{mfss}
\langle m \rangle = L^{- \beta/\nu} {\cal F}_1 \left( \tau L^{1/\nu}, R L^{r} \right),
\end{equation}
and Eq.~(\ref{m2scaling}) by
\begin{eqnarray}
{\cal F}(X,Y)&=&Y^{\beta/\nu}{\cal F}_1\left(XY^{-1/\nu},Y^{-r}\right),\nonumber\\%\label{cf1}\\
F(X,Y)&=&Y^{2\beta/\nu}F_1\left(XY^{-1/\nu},Y^{-r}\right),\label{f3}
\end{eqnarray}
respectively, similar to Eq.~(\ref{f12}). Crossovers in $\langle m\rangle$ and $\langle m^2\rangle$ similar to Eq.~(\ref{ftsrl}) for $\chi$ would then result from Eqs.~(\ref{m2scaling}) and (\ref{mftsr}) to (\ref{f3}).

However, we need a new ingredient. There is an important difference between heating and cooling. During cooling, because of the absence of a symmetry breaking direction, the average magnetization and its squared must be vanished in the thermodynamic limit, though their appropriate difference, $\chi$, does not near $T_c$. In FSS, the volume factor for the average just leads to the proper power of the size factor as has been seen above. In FTS, on the other hand, the effective length scale due to the driving---which is shorter than the system size---divides, as may be envisioned, the system into dynamically fluctuating regions of correlation. The average of the magnetization and its squared of these regions ought to be vanished in the thermodynamic limit owing to the central limit theorem. In heating, however, there exist finite averaged magnetization and its squared. These then imply that for cooling in the FTS regime, the scaling function ${\cal F}_0(Y)$ behaves for small $Y$ as
\begin{equation}
{\cal F}_0(Y)=Y^{d/2}{\cal \hat{F}}_0(Y)\quad {\rm for~} Y=L^{-1}R^{-1/r}\ll 1 \label{sing}
\end{equation}
with the caret indicating an associated regular scaling function for finite $R$. So does $F_{0}$ for cooling. Equation~(\ref{sing}) reflects the restriction to the fluctuations in cooling, since the expansion of ${\cal F}_0(Y)$ for small $Y$ must now start from order $d/2$ instead of $0$. Accordingly, for cooling, the FTS and FSS forms and their crossover are
\begin{eqnarray}
  \langle m\rangle &=& \left \{
\begin{array}{l l}
L^{-\beta/\nu} {\cal F}_{10} \left(R L^{r} \right),  &  R < R_{\rm KZ}(L),\\
& \\
L^{-d/2}R^{-\gamma/2r\nu} {\cal \hat{F}}_{0}\left(L^{-1}R^{-1/r}\right), & R > R_{\rm KZ}(L),
\end{array}
\right.\nonumber\\
\label{mftsrlc}\\
  \langle m^2\rangle &=& \left \{
\begin{array}{l l}
L^{-2\beta/\nu} F_{10} \left(R L^{r} \right),  &  \quad~ R < R_{\rm KZ}(L),\\
& \\
L^{-d}R^{-\gamma/r\nu} \hat{F}_{0}\left(L^{-1}R^{-1/r}\right),  & \quad~ R > R_{\rm KZ}(L),
\end{array}
\right.\nonumber\\
\label{m2ftsrlc}
\end{eqnarray}
while for heating,
\begin{eqnarray}
  \langle m\rangle &=& \left \{
\begin{array}{l l}
L^{-\beta/\nu} {\cal F}_{10} \left(R L^{r} \right),  &  ~R < R_{\rm KZ}(L)\\
& \\
R^{\beta/r\nu} {\cal F}_{0}\left(L^{-1}R^{-1/r}\right), & ~R > R_{\rm KZ}(L),
\end{array}
\right.\label{mftsrlh}\\
\nonumber\\
  \langle m^2\rangle &=& \left \{
\begin{array}{l l}
L^{-2\beta/\nu} F_{10} \left(R L^{r} \right),  &  R < R_{\rm KZ}(L)\\
& \\
R^{2\beta/r\nu} F_{0}\left(L^{-1}R^{-1/r}\right),  & R > R_{\rm KZ}(L),
\end{array}
\right.\label{m2ftsrlh}
\end{eqnarray}
where we have used Eq.~(\ref{dn}) to combine the two exponents of $R$ into one. One sees therefore that the second line of Eq.~(\ref{m2ftsrlc}) is just the complete form of Eq.~(\ref{m2vx}), the LPS scaling, which leads to a vanishing $\langle m^2\rangle$ in the thermodynamic limit as expected. This is to be compared with the heating case where $\langle m^2\rangle$ remains finite in the FTS regime $R > R_{\rm KZ}(L)$ in the same limit. Owing to this difference, opposite signs of the powers of the leading dependences on $R$ for cooling and heating occur. Therefore, we have derived directly the LPS scaling.

Moreover, from the derivation, we see that the LPS scaling~(\ref{m2vx}) is a consequence of the FTS and the central limit theorem. It may thus be regarded as an interplay of FTS and FSS and be termed \emph{finite-time-finite-size scaling}, though a trivial form of FSS in the sense that only the dimension of the space matters. However, from Eqs.~(\ref{mftsrlc}) to (\ref{m2ftsrlh}) and Eq.~(\ref{ftsrl}), one sees, on the other hand, that this LPS' scaling is only true for some observables in cooling. For heating and for $\chi$ in cooling for example, no $L^d$-like factor appears as the finite magnetization removes the restriction. In addition, one sees from Eqs.~(\ref{mftsrlc}) to (\ref{m2ftsrlh}) that although the FTS forms are different in heating and cooling, the FSS forms are identical.

Because of Eqs.~(\ref{chi}) and (\ref{dn}), the relations among the magnetization and its squared and the susceptibility, there must exist relationship between the three sets of scaling functions $f$, $F$, and ${\cal F}$. Indeed, Eqs.~(\ref{mftsrlc}) and (\ref{m2ftsrlc}) along with Eqs.~(\ref{chi}) and (\ref{dn}) lead to Eq.~(\ref{ftsrl}) with $f_{0}=\hat{F}_{0}-{\cal \hat{F}}_0^2$ and $f_{10}=F_{10}-{\cal F}_{10}^2$; while the second lines of Eqs.~(\ref{mftsrlh}) and (\ref{m2ftsrlh}) are consistent with the second line of Eq.~(\ref{ftsrl}) if $f_{0}(Y)=Y^{-d}[F_{0}(Y)-{\cal F}_0^2(Y)]$. The singularity of $Y^{-d}=L^dR^{d/r}$ for large $L$ is just canceled by $Y^d$ arising from the difference of $F_{0}$ and ${\cal F}_0^2$, which themselves are not singular and thus no $L^{-d}$ factors appear in Eqs.~(\ref{mftsrlh}) and (\ref{m2ftsrlh}). This is the reason why both the \emph{standard} (without the $L^{-d}$ factor) FTS forms of $\langle m^2\rangle$ or $\langle m\rangle$ and $\chi$ are correct in heating, whereas only the standard FTS form of $\chi$ is correct in cooling in the FTS regime.

In order to further show the limitation of the LPS scaling, we apply a small external field $H$ to the system in cooling. The FTS form for $\langle m^2\rangle$ becomes~\cite{Gong10,Zhong11}
\begin{equation}
\langle m^2\rangle = R^{2\beta/r\nu}F_H\left(\tau R^{-1/r\nu},HR^{-\beta\delta/r\nu}, L^{-1}R^{-1/r}\right),\label{mftsh}
\end{equation}
where $\delta$ is another critical exponent. This small applied field results in a finite magnetization if it is sufficiently large. In this case, the scaling function $F_{H0}$ at $\tau=0$ ought to be regular and the cooling $\langle m^2\rangle$ now behaves as in heating for fixed $HR^{-\beta\delta/r\nu}$. Similar results apply to $\langle m\rangle$ too.

In addition, we have clearly seen that LPS' adiabatic and universal scaling regimes are just the FSS and FTS regimes, respectively. Consequently, they are both manifestly universal scaling regimes. We shall see below that when the externally applied field is not large enough, there appears between the standard FTS and FSS regimes an intermediate regime in which the LPS scaling holds. We thus call this regime as the finite-time-finite-size scaling regime instead of the universal scaling regime to distinguish it from the standard FTS regime.

Finally, we consider the scaling function $F_{30}$ for large $R$ in Eq.~(\ref{2scaling_functions}). As pointed out in Sec.~\ref{kz}, this scaling function cannot in fact describe the data for large $R$ since they all deviate from the scaling function as can be seen from Fig.~4 for $\langle m^2 \rangle L^d$ versus $R^{-1}$ in Ref.~[\onlinecite{Liu}]. So, it is at best another presentation of the universal scaling regime~\cite{Liu}. This description can be understood from Eq.~(\ref{m2ftsrlc}) or Eq.~(\ref{ftsrl}): It is just an approximation representation of the scaling function in the FTS regime. Indeed, the leading contribution to $L^d\langle m^2\rangle$ or $\chi$ from $R^{-1}$ is an exponent $\gamma/r\nu$, which is just $x$ from Eq.~(\ref{xgamma}) in agreement with the results in Ref.~[\onlinecite{Liu}]. However, we find (not shown) that even presented in $L^d\langle m^2\rangle$ or $\chi R^{\gamma/r\nu}$ versus $L^{-1}R^{-1/r}$ for FTS, the data for large $R$ do not collapse onto a single curve. This means that even higher-order expansion of $L^{-1}R^{-1/r}$ cannot account for the deviations. As $R$ is rather large and the state of the system is hardly changed during the cooling, these data of large $R$ are far away from the critical point. Either they may exhibit no scaling at all or at least corrections to scaling~\cite{Wegner} have to be considered. In our simulations below, we choose relatively small rates as our results will show and shall not consider them further.

\section{\label{chara}Characteristics of FSS and FTS graphs}
In this section, we summarize the main characteristics of graphs for $\langle m\rangle$, $\langle m^2\rangle$,
and $\chi$ when presented in the forms of FSS and FTS for ease of comparison with numerical results.
By the form of FSS (FTS), we mean the graphs are rescaled by appropriate powers of $L$ ($R$) and thus in FSS (FTS).

Before we present the characteristics, we note that in Eqs.~(\ref{ftsrl}) and (\ref{mftsrlc}) to (\ref{m2ftsrlh}), the scaling functions on the right hand sides are all regular when their arguments are small and thus lie in the regimes specified. Accordingly, they can all be expanded and consist of sub-leading contributions to the leading features. In the following summary, we ignore them temporally and only focus on the leading features.

Equations~(\ref{ftsrl}) and (\ref{f12}) imply that when presented in the FSS plane of $\chi L^{-\gamma/\nu}$ versus $RL^r$ on a double-logarithmic scale, i.e., in the form of FSS, the leading FSS regime is a horizontal line, while the leading FTS regime is a line with a slope $-x=-\gamma/r\nu$. On the other hand, when presented in the FTS plane of $\chi R^{\gamma/r\nu}$ versus $L^{-1}R^{-1/r}$ on a double-logarithmic scale, viz., in the form of FTS, the leading FTS regime is a horizontal line, while the leading FSS regime is a line with a slope $-\gamma/\nu$. These apply to both heating and cooling.

For cooling, in the form of FSS, Eqs.~(\ref{mftsrlc}) and (\ref{m2ftsrlc}) together with Eqs.~(\ref{f3}) and (\ref{sing}) imply that $\langle m\rangle L^{\beta/\nu}$ ($\langle m^2\rangle L^{2\beta/\nu}$) versus $L^rR$ on a double-logarithmic scale is a horizontal line in the leading FSS regime, while it is a line with a slope of $-\gamma/2r\nu$ ($-\gamma/r\nu$) in the leading FTS regime. On the other hand, in the form of FTS, $\langle m\rangle R^{-\beta/r\nu}$ ($\langle m^2\rangle R^{-2\beta/r\nu}$) versus $L^{-1}R^{-1/r}$ on a double-logarithmic scale is a line with a slope of $d/2$ ($d$) in the leading FTS regime, while its slope changes to $\beta/\nu$ ($2\beta/\nu$) in the leading FSS regime. In addition, one can also presented in the form of FTS, or more specifically, finite-time-finite-size scaling, $\langle m\rangle L^{d/2}R^{\gamma/2r\nu}$ ($\langle m^2\rangle L^{d}R^{\gamma/r\nu}$) versus $L^{-1}R^{-1/r}$ on a double-logarithmic scale. In this case, the leading FTS regime is a horizontal line, while the leading FSS regime now has a slope of $-\gamma/2\nu$ ($-\gamma/\nu$) according to Eqs.~(\ref{f3}) to (\ref{m2ftsrlc}).

For heating, from Eqs.~(\ref{mftsrlh}), (\ref{m2ftsrlh}), and Eq.~(\ref{f3}), in the FSS form, the horizontal line of $\langle m\rangle L^{\beta/\nu}$ ($\langle m^2\rangle L^{2\beta/\nu}$) versus $L^rR$ on a double-logarithmic scale in the leading FSS regime now crosses over smoothly to an oblique line of a slope $\beta/r\nu$ ($2\beta/r\nu$) in the leading FTS regime, different from cooling. On the other hand, in the FTS form, $\langle m\rangle R^{-\beta/r\nu}$ ($\langle m^2\rangle R^{-2\beta/r\nu}$) versus $L^{-1}R^{-1/r}$ on a double-logarithmic scale is now a horizontal line in the leading FTS regime and crosses over to an oblique line with again a slope of $\beta/\nu$ ($2\beta/\nu$) in the leading FSS regime. In addition, one can again presented in the form of finite-time-finite-size $\langle m\rangle L^{d/2}R^{\gamma/2r\nu}$ ($\langle m^2\rangle L^{d}R^{\gamma/r\nu}$) versus $L^{-1}R^{-1/r}$ on a double-logarithmic scale. In this case, the leading FTS and FSS regimes now have a slope of $-d/2$ ($-d$) and $-\gamma/2\nu$ ($-\gamma/\nu$), respectively, according to Eqs.~(\ref{mftsrlh}), (\ref{m2ftsrlh}), (\ref{f3}), and (\ref{dn}). All these results are summarized in Table~\ref{charac}, where we give the slope of each regime in a graph, represented by its ordinate and abscissa, on a double-logarithmic scale. Note that the FTS and FSS in the title of the table represent the form of a graph, whereas they stand for the leading regimes in the graph.
\begin{table}
\caption{\label{charac} Leading characteristics of graphs in FTS and FSS.}
\begin{ruledtabular}
\begin{tabular}{llcccc}
&&\multicolumn{2}{c}{heating}&\multicolumn{2}{c}{cooling}\\
ordinate& abscissa & FTS & FSS & FTS & FSS\\
\hline
$\langle m^2\rangle L^{2\beta/\nu}$  &$RL^{r}$        &$2\beta/r\nu$  & 0             & $-\gamma/r\nu$  &0\\
$\langle m^2\rangle R^{-2\beta/r\nu}$&$L^{-1}R^{-1/r}$&0              & $2\beta/\nu$  & $d$             &$2\beta/\nu$\\
$\langle m\rangle L^{\beta/\nu}$     &$RL^{r}$        &$\beta/r\nu$   & 0             & $-\gamma/2r\nu$ &0\\
$\langle m\rangle R^{-\beta/r\nu}$   &$L^{-1}R^{-1/r}$&0              & $\beta/\nu$   & $d/2$           &$\beta/\nu$\\
$\chi L^{-\gamma/\nu}$               &$RL^{r}$        &$-\gamma/r\nu$ & 0             & $-\gamma/r\nu$  &0\\
$\chi R^{\gamma/r\nu}$               &$L^{-1}R^{-1/r}$&0              & $-\gamma/\nu$ & 0               & $-\gamma/\nu$\\
$\langle m^2\rangle L^{d}R^{\gamma/r\nu} $&$L^{-1}R^{-1/r}$&$-d$      & $-\gamma/\nu$ & 0               & $-\gamma/\nu$\\
$\langle m\rangle L^{d/2}R^{\gamma/2r\nu}$&$L^{-1}R^{-1/r}$&$-d/2$    & $-\gamma/2\nu$& 0               & $-\gamma/2\nu$\\
\end{tabular}
\end{ruledtabular}
\end{table}

These characteristics offers a possible method to estimate the critical exponents and the division of the FTS and FSS regimes when the sub-leading contributions are taken into account. As an example, assume that $T_c$ and the static critical exponents are known and we want to estimate the dynamic critical exponent $z$, as was done for the 2D and 3D Ising models~\cite{Zhong06,Liu}. Using Eq.~(\ref{m2ftsrlh}) as an example, we can write it as
\begin{eqnarray}
  \langle m^2\rangle R^{-2\beta/r\nu} &=& \left \{
\begin{array}{l l}
Y^{2\beta/\nu}F_{10}\left(Y^{-1/r}\right) ,  &  \qquad\quad ~Y > 1\\
\\
F_{0}(Y),  &\qquad\quad ~Y < 1,
\end{array}
\right.\nonumber\\
&\approx& \left\{
\begin{array}{l l}
Y^{2\beta/\nu}\sum\limits_{k=0}^{n_{k}}\frac{1}{k!}F_{10}^{(k)}(0)Y^{-k/r},  &  Y > 1,\\
\sum\limits_{k=0}^{n_{k}'}\frac{1}{k!}F_{0}^{(k)}(0)Y^k,  & Y < 1,
\end{array}
\right.\nonumber\\
\label{m2ftsrlhs}
\end{eqnarray}
in the FTS form, while it becomes
\begin{eqnarray}
  \langle m^2\rangle L^{2\beta/\nu} &=& \left \{
\begin{array}{l l}
F_{10} (\check{Y}),  &  \qquad\quad ~~\check{Y} < 1\\
\\
\check{Y}^{2\beta/r\nu} F_{0}\left(\check{Y}^{-1/r}\right),  &\qquad\quad ~~\check{Y} > 1,
\end{array}
\right.\nonumber\\
&\approx& \left\{
\begin{array}{l l}
\sum\limits_{k=0}^{n_{k}''}\frac{1}{k!}F_{10}^{(k)}(0)\check{Y}^k,  &  \check{Y} < 1,\\
\check{Y}^{2\beta/r\nu}\sum\limits_{k=0}^{n_{k}'''}\frac{1}{k!}F_{0}^{(k)}(0)\check{Y}^{-k/r},  & \check{Y} > 1,
\end{array}
\right.\nonumber\\
\label{m2ftsrlhss}
\end{eqnarray}
in the FSS form, where $\check{Y}=Y^{-r}=RL^r$. So, we can find the optimal $z$ from the least $\chi^2$ (not to be confused with the susceptibility) per degree of freedom (dof) by varying both the division of the FSS and FTS regimes and the maximal degrees $n_k$ and $n_{k}'$ or $n_k''$ and $n_k'''$ and fitting $\langle m^2\rangle$ at $T_c$ in heating to the two expressions of Eq.~(\ref{m2ftsrlhs}) in FTS or Eq.~(\ref{m2ftsrlhss}) in FSS for the two regimes, respectively. Note that the division found in this way is only an approximate representative as it is method dependent.

In addition to this ``two-side'' method, ``one-side'' method also works in which we just fit the FTS regime in each graph by expanding the scaling function as in Eqs.~(\ref{m2ftsrlhs}) and (\ref{m2ftsrlhss}). However, in this case, the range of the FTS regime within which the least $\chi^2/{\rm dof}$ is obtained may well be different from the division between the FTS and FSS regimes found by the two-side method.

\section{\label{mm}Model and methods}
In order to test our results, we consider the classical Ising model defined by the Hamiltonian
\begin{equation}
\label{eq:Hising}
H = -J \sum_{\left\langle i,j \right\rangle} \sigma_i \sigma_j,
\end{equation}
where $J > 0$ is a coupling constant, $\sigma_i = \pm 1$ is a spin at site $i$, and the summation is over all nearest neighbor pairs. Periodic boundary conditions are applied throughout. The order parameter and its square are defined as
\begin{equation}
\label{eq:m2def}
m = \frac{1}{N} \sum_{i=1}^N \sigma_i,\qquad m^2 =\left ( \frac{1}{N} \sum_{i=1}^N \sigma_i  \right )^2
\end{equation}
for the $N$ spins.
For cooling, since the magnetization $\langle m\rangle$ vanishes when the system is cooled linearly to $T_c$, we use $\langle |m|\rangle$ instead. When $L$ and $R$ are small, there is also appreciable possibility for $m$ to flip. So, we use $\langle |m|\rangle$ for heating too. We find that this choice affects the width of the crossover region between the FTS and FSS regimes. Accordingly, the susceptibility changes to
\begin{equation}\label{chimm}
\chi'=L^d\left(\langle m^2\rangle-\langle |m|\rangle^2\right).
\end{equation}
$\chi'$ approaches $(1-2/\pi)\chi$ in the disorder state but diverges with the same critical exponent~\cite{Binder}, which we have confirmed. So, we use $\langle |m|\rangle$ and the definition~(\ref{chimm}) without the absolute symbols and the prime hereafter.

We shall study the model on 2D square and 3D simple cubic lattices. For the 2D square lattice, the exact value of $T_c$ and the critical exponents are known exactly~\cite{Ferrenberg,Pelissetto,Kleinert}:
$T_c = 2/\ln(1+\sqrt{2}) J$, $\nu=1$, $\beta=1/8$, $\delta=15$, and $\gamma=7/4$. For the 3D cubic lattice, the value of $T_c$ and the critical exponents have been estimated to rather high precisions~\cite{Ferrenberg,Pelissetto,Kleinert}: $T_c$=1/0.221~659~5(26), $\nu=0.630~1(4)$, $\beta=0.326~5(3)$, and $\gamma=1.237~2(5)$. For the dynamic critical exponent $z$, much work has been devoted to extracting it for the 2D and 3D Ising models using Metropolis dynamics. The values obtained in 2D are typically close to 2.2, with $z=2.1667(5)$, obtained in Ref.~[\onlinecite{Nightingale}]. The average of the values in 3D is $z=2.031(3)$, estimated in Refs.~[\onlinecite{Grassberger}] and [\onlinecite{Kikuchi}]. Using these extant exponents, we list in Table~\ref{numc} the numerical values of some entries in Table~\ref{charac}.
\begin{table}
\caption{\label{numc} Extant values of some slopes.}
\begin{ruledtabular}
\begin{tabular}{lcccc}
   & $\beta/\nu$ & $\beta/r\nu$ & $\gamma/\nu$ & $\gamma/r\nu$\\
\hline
2D & 1/8         & 0.039473(6)    & 7/4          & 0.55263(9)\\
3D & 0.5182(6)   & 0.1432(3)      & 1.964(2)     & 0.5427(7)\\
\end{tabular}
\end{ruledtabular}
\end{table}

We use the single-spin Metropolis algorithm \cite{Metropolis} for the dynamics with the time unit given by $N$ random attempts to flip the spins.
The system starts at an initial temperature $T_i$ far away from the critical point, with the value of critical temperature $T_c$ quoted above. Usually, the distance between $T_i$ and $T_c$ is chosen to be about $0.5T_c$ with an integer multiple of $R$. Thus, the system can land at the critical point $T_c$ within a certain step, when it is cooled or heated linearly after equilibration. All the results are measured exactly at the point satisfied $\tau=0$, that is $T=T_c$.  Typically, we calculated averages over between 5 and 10 thousand samples for each rate $R$ on each lattice size $L$.

\section{\label{results}Simulation results}
In this section, we shall first display the scaling collapses in the FTS and FSS graphs to show the leading FTS and FSS features and their crossovers in Sec.~\ref{ftsfss}. The sharp difference between heating and cooling according to the theory will be confirmed. This will affect the critical exponents estimated with FTS as will be illustrated in Sec.~\ref{eof}. Finally, we present our results for the dynamic exponent $z$ estimated from the graphs in Sec.~\ref{zvalues}.

\subsection{\label{ftsfss}FTS and FSS and their crossover}
\begin{figure}
  \centering
  \includegraphics[width=\columnwidth,clip=true]{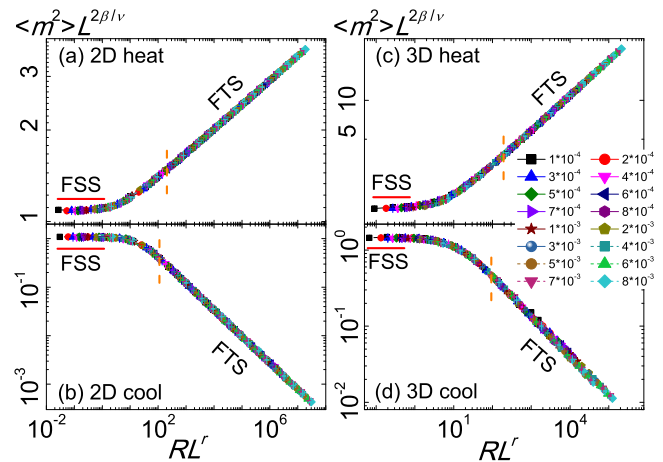}
  \caption{(Color online) FSS collapses of the squared magnetization at $T_c$ for linear heating and cooling in 2D and 3D. Error bars for the data are shown but are smaller than the symbol sizes. The legend gives the rate $R$ used for both heating and cooling and 2D and 3D and applies to all FSS graphs. The (red) line segments near FSS in this and all following figures indicate the slopes of the leading FSS regimes according to the extant values listed in Table~\ref{numc}. Similarly, the vertical dashed lines in all figures mark the divisions of the FTS and FSS regimes fitted with the two-side method (see the text for details). The $z$ values in all figures are given in Table~\ref{tablez} below except explicitly given otherwise. Note that the slopes of the leading FTS sections are different for cooling and heating, with the values of (a) $0.0810(2)$ and (c) $0.2718(5)$ in agreement with $2\beta/r\nu$ for heating according to Eq.~(\ref{m2ftsrlh}) and Table~\ref{numc} and (b) $-0.5478(7)$ and (d) $-0.532(3)$ in agreement with $-\gamma/r\nu$ for cooling according to Eq.~(\ref{m2ftsrlc}) and Table~\ref{numc}, respectively.}
   \label{m2L^}
\end{figure}
\begin{figure}
  \centering
  \includegraphics[width=\columnwidth,clip=true]{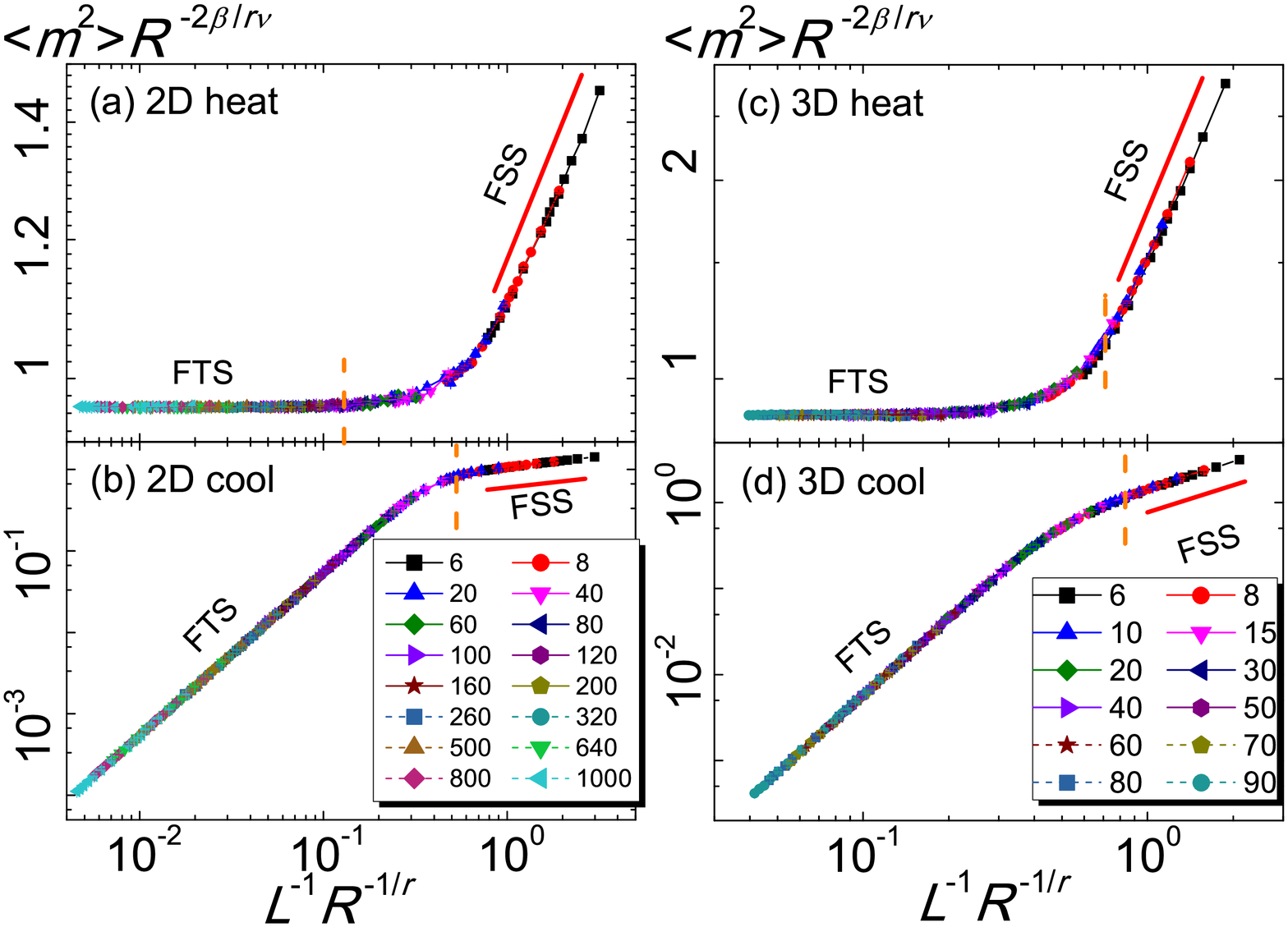}
  \caption{(Color online) FTS collapses of the same data as in Fig.~\ref{m2L^}. Lines connecting data are guides to the eye in all figures. The legends gives the sizes of the lattices used and applies to all FTS graphs except otherwise given explicitly. The (red) line segments and the vertical dashed lines share the same meaning as in Fig.~\ref{m2L^}. For heating, the horizontal sections are the leading FTS regimes and the oblique sections parallel to the lines with the slopes of $2\beta/\nu$ given in Table~\ref{numc} are the leading FSS regimes. For cooling, the oblique sections with slopes close to $d$ are the leading FTS regimes and the other oblique sections are the leading FSS regimes with similar slopes to the heating case.}
   \label{m2R^}
\end{figure}
\begin{figure}
  \centering
  \includegraphics[width=\columnwidth,clip=true]{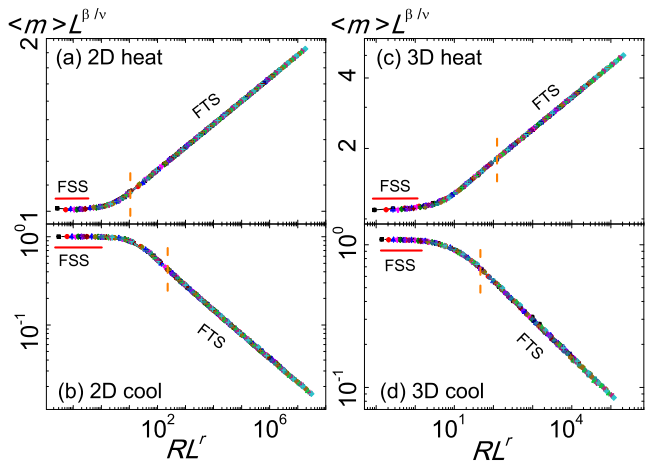}
  \caption{(Color online) FSS collapses of the magnetization at $T_c$ for linear heating and cooling in 2D and 3D. All symbols share the same meaning as in Fig.~\ref{m2L^}. Similar to Fig.~\ref{m2L^}, the slopes of oblique leading FTS sections are different for heating and cooling, with the values of (a) $0.0405(2)$ and (c) $0.1348(5)$, in agreement with $\beta/r\nu$ according to Eq.~(\ref{mftsrlh}) and Table~\ref{numc} and (b) $-0.2807(8)$ and (d) $-0.271(2)$, in agreement with $-\gamma/2r\nu$ according to Eq.~(\ref{mftsrlc}) and Table~\ref{numc}, respectively.}
   \label{mL^}
\end{figure}
\begin{figure}
  \centering
  \includegraphics[width=\columnwidth,clip=true]{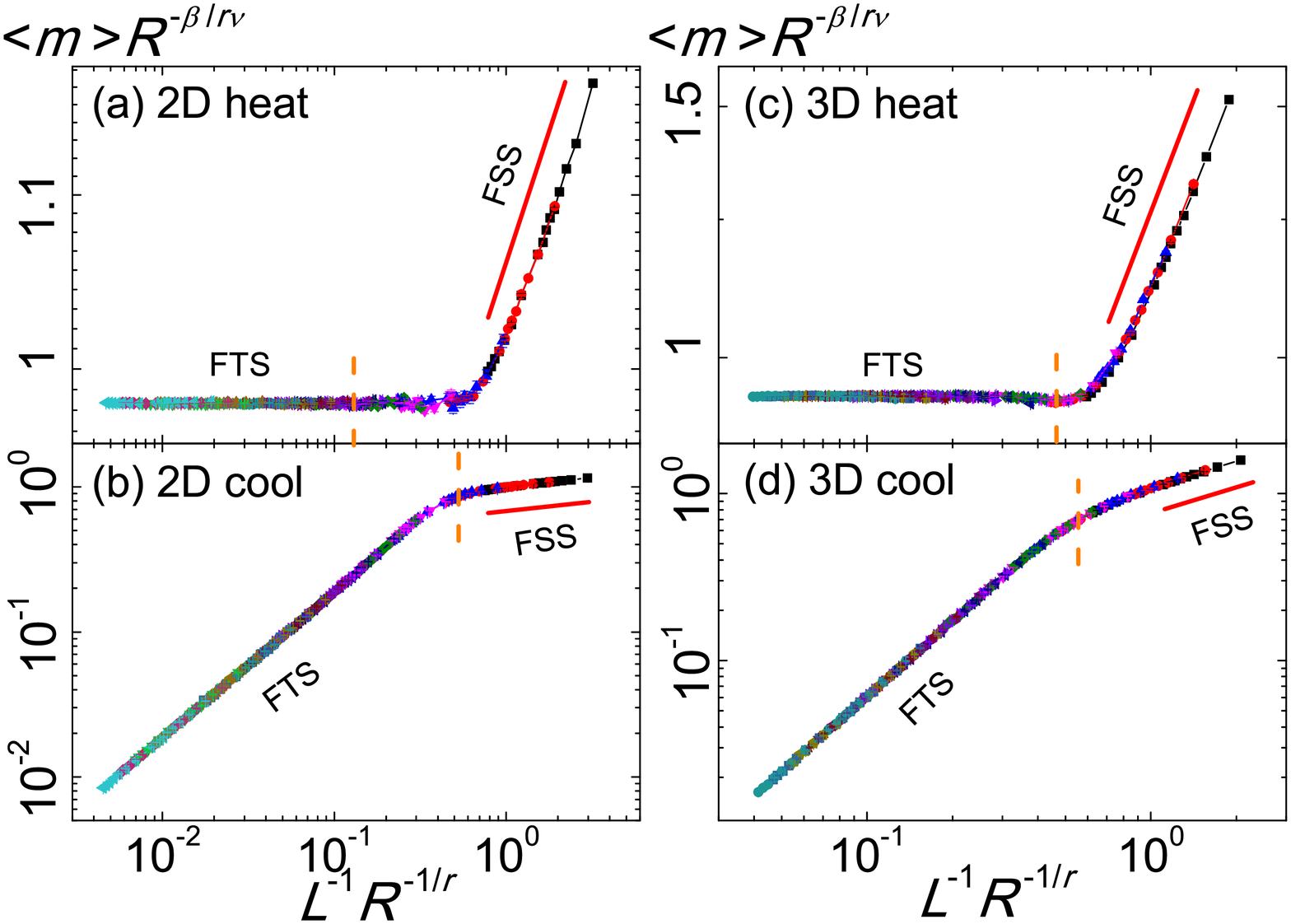}
    \caption{(Color online) FTS collapses of the same data as in Fig.~\ref{mL^}. All symbols share the same meaning as in Fig.~\ref{m2R^}. For heating, the horizontal sections are the leading FTS regimes and the oblique sections parallel to the lines with the slopes of $\beta/\nu$ given in Table~\ref{numc} are the leading FSS regimes. For cooling, the oblique sections with slopes close to $d/2$ are the leading FTS regimes and the other oblique sections are the leading FSS regimes with similar slopes to the heating case.}
   \label{mR^}
\end{figure}
\begin{figure}
  \centering
  \includegraphics[width=\columnwidth,clip=true]{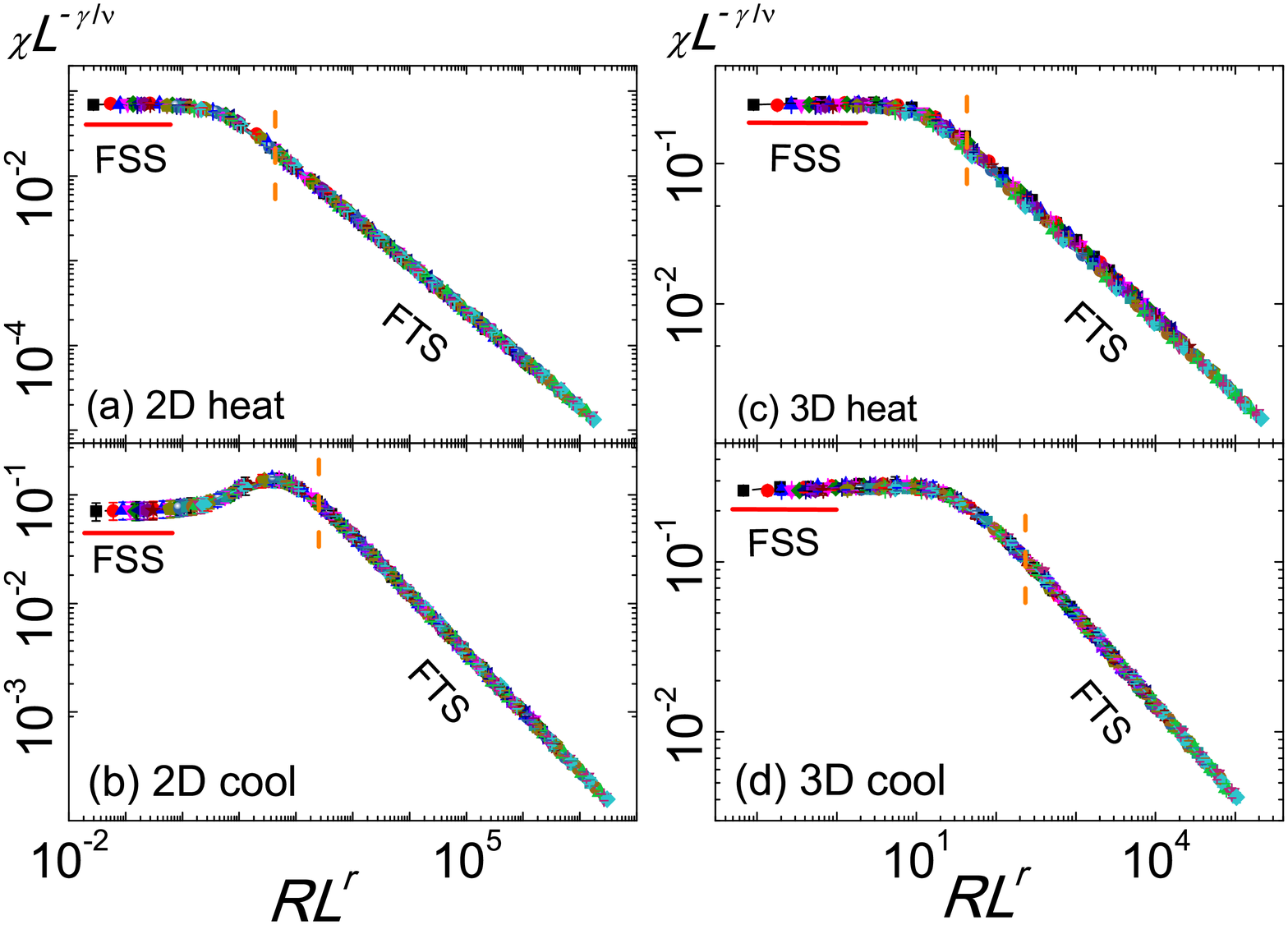}
  \caption{(Color online) FSS collapses of the susceptibility at $T_c$ for linear heating and cooling in 2D and 3D. Some error bars are visible here. Due to the large fluctuations, the vertical dashed lines are estimated by fixing both $z$ to the average (column 5, Table~\ref{tablez} below) and the maximal degrees of the expansions in Eq.~(\ref{m2ftsrlhss}). The leading FTS regimes have slopes of $-0.562(1)$, $-0.540(1)$, $-0.557(4)$, and $-0.526(4)$ from (a) to (d), respectively, in agreement with $-x=-\gamma/r\nu$ listed in Table~\ref{numc} for 2D and 3D, whereas the leading FSS regimes are all horizontal, in consistence with Eq.~(\ref{ftsrl}) and Table~\ref{charac}.}
   \label{chiL^}
\end{figure}
\begin{figure}
  \includegraphics[width=\columnwidth,clip=true]{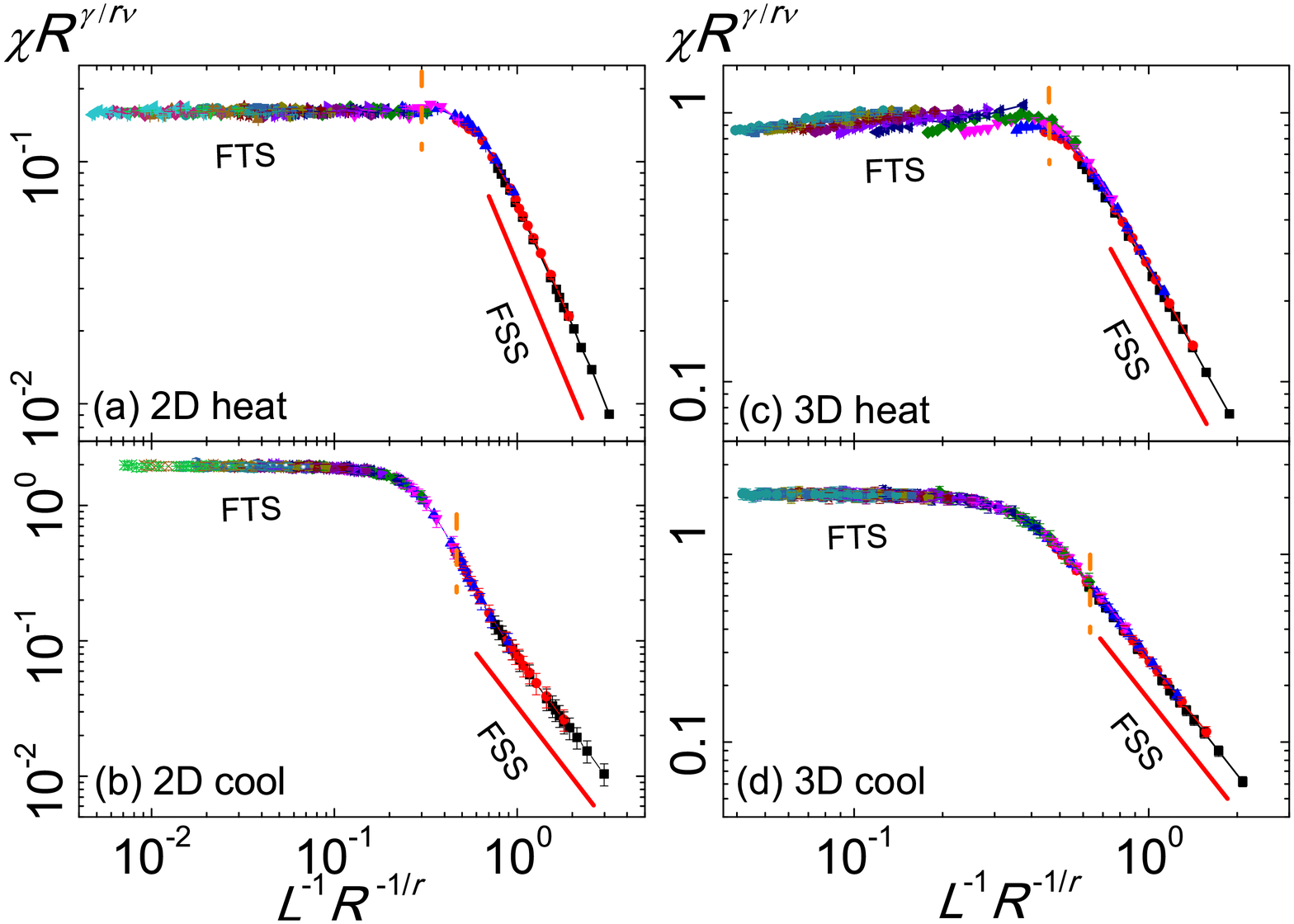}
  \caption{(Color online) FTS collapses of the same data as in Fig.~\ref{chiL^}. The horizontal sections now correspond to the leading FTS regimes, while the oblique sections parallel to the red line segments with the slopes $-\gamma/\nu$ given in Table~\ref{numc} are the leading FSS regimes, in agreement with Eq.~(\ref{ftsrl}) and Table~\ref{charac}. The vertical dashed lines that separate the FTS and FSS regimes are estimated with a method similar to that in Fig.~\ref{chiL^}.}
   \label{chiR^}
\end{figure}
Figures~\ref{m2L^} to \ref{chiR^} show the FSS and FTS collapses of $\langle m^2\rangle$, $\langle m\rangle$, and $\chi$ at $T_c$ in 2D and 3D. The results from both heating and cooling are shown. In each figure, the data collapse well onto a single curve after rescaling. Each rescaled curve consists of a leading FTS section and a leading FSS section with different slopes and a crossover between them. However, all the three sections are describable by Eqs.~(\ref{ftsrl}) and (\ref{mftsrlc}) to (\ref{m2ftsrlh}). So, we can employ the two-side method proposed in Sec.~\ref{chara} to approximately divide the FTS regime and the FSS regime. The results are shown as vertical dashed lines in the figures. One sees that because of the sub-leading contribution, the FTS regime, which consists of the LPS universal scaling regime, is not necessarily only straight. Note that, in the figures, we refer to the leading feature (the straight section) of the FTS (FSS) regime as the leading FTS (FSS) regime, although the FTS (FSS) regime also includes the curved crossover section. From Figs.~\ref{m2L^} to \ref{chiR^}, it is seen that while $\chi$ behaves similarly in heating and cooling, $\langle m^2\rangle$ and $\langle m\rangle$ exhibit distinct features in general and sharp contrast in the leading FTS regimes in heating and cooling in Figs.~\ref{m2L^} and \ref{mL^} in particular, in agreement with the leading characteristics listed in Table~\ref{charac} and thus with the theory, Eqs.~(\ref{ftsrl}) and (\ref{mftsrlc}) to (\ref{m2ftsrlh}). This is different from behaviors of the leading FSS regimes, which show identical slopes in both heating and cooling for each observable.

We note that in Fig.~\ref{chiR^}(c) for the 3D $\chi R^{\gamma/r\nu}$ in heating, the collapse in the horizontal FTS regime appear not so good. In fact, the same regime in the FSS form, which is oblique, Fig.~\ref{chiL^}(c), may seem somewhat rough in comparison with others. This indicates that FTS is more sensitive than FSS. We do not yet know why the 3D $\chi$ in heating is poor, since all other observables in 3D in heating appear quite good. A possible reason may be our range of lattice sizes in 3D is not large, which may affect the accuracy of $z$ estimated. Another reason may be the fluctuations of $\chi$ are large at $T_c$. Nevertheless, the overall feature still agrees with the theory.

\begin{figure}
  \centering
  \includegraphics[width=\columnwidth,clip=true]{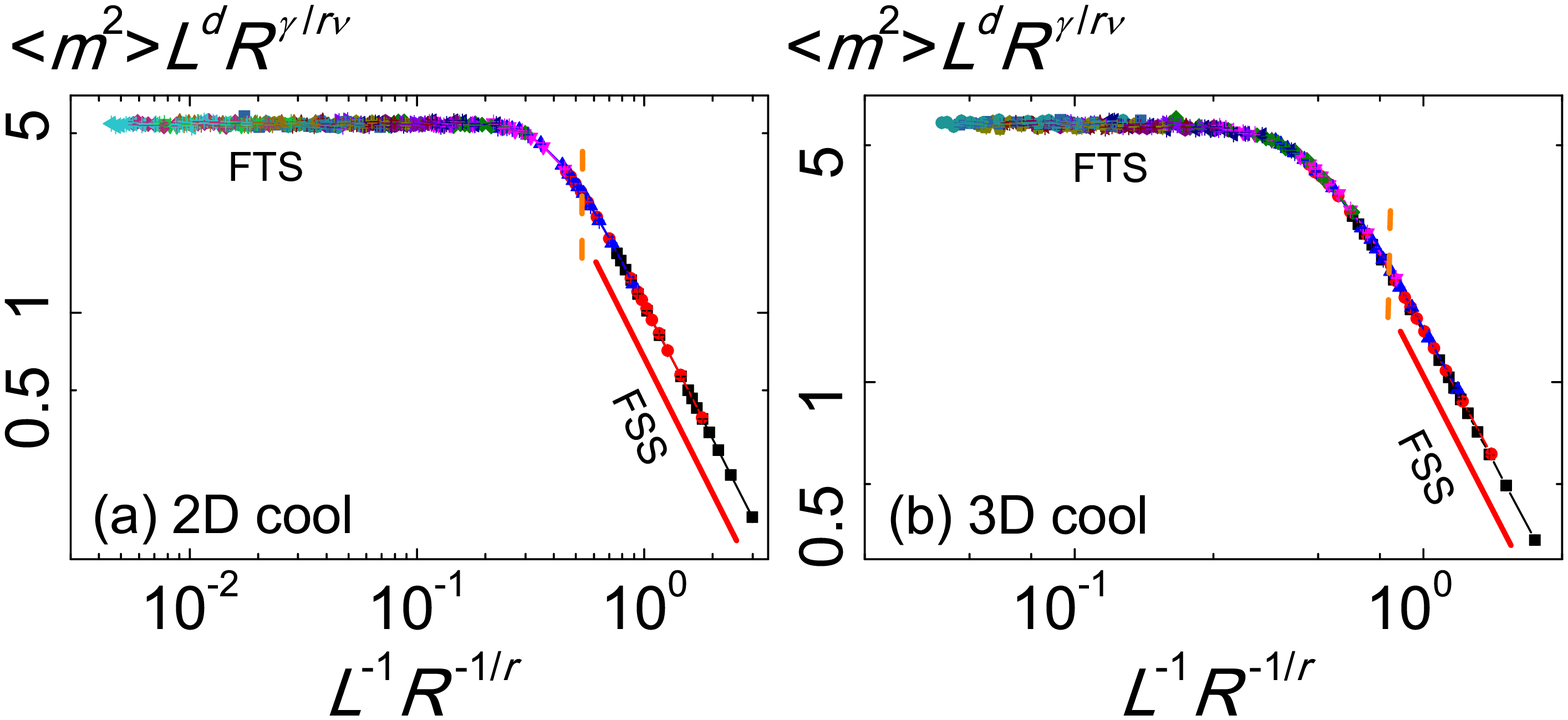}
  \vskip -2cm
  \caption{(Color online) \label{downm2R^} Finite-time-finite-size scaling collapses of the same data as in Fig.~\ref{m2L^}(b) and (d) and Fig.~\ref{m2R^}(b) and (d) for cooling. The leading FTS sections now become horizontal rather than oblique as in Figs.~\ref{m2L^}(b) and (d) and \ref{m2R^}(b) and (d), while the slopes of the leading FSS regimes change to $-\gamma/\nu$ given in Table~\ref{numc} and illustrated by the red line segments, in agreement with Eq.~(\ref{m2ftsrlc}) and Table~\ref{charac}.}
\end{figure}
\begin{figure}
  \centering
  \includegraphics[width=\columnwidth,clip=true]{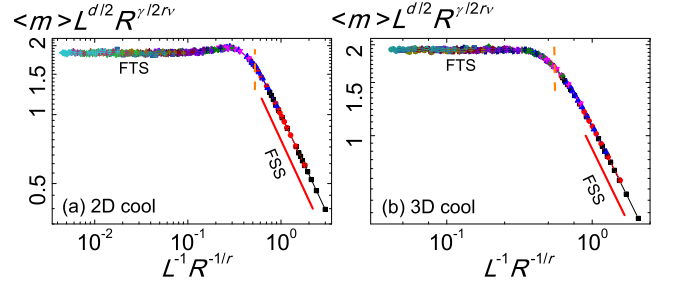}
  \vskip -2cm
  \caption{(Color online) \label{downmR^} Finite-time-finite-size scaling collapses of the same data as in Fig.~\ref{mL^}(b) and (d) and Fig.~\ref{mR^}(b) and (d) for cooling. Similar to Fig.~\ref{downm2R^}, the leading FTS sections now become horizontal rather than oblique as in Figs.~\ref{mL^}(b) and (d) and \ref{mR^}(b) and (d), while the slopes of the leading FSS regimes change to $-\gamma/2\nu$ given in Table~\ref{numc} and illustrated by the red line segments, in agreement with Eq.~(\ref{mftsrlc}) and Table~\ref{charac}.}
\end{figure}
To confirm the LPS finite-time-finite-size scaling in cooling, we plot in Figs.~\ref{downm2R^} and \ref{downmR^} $\langle m^2\rangle L^{d}R^{-\gamma/r\nu}$ and $\langle m\rangle L^{d/2}R^{-\gamma/2r\nu}$, respectively, for cooling. According to Eqs.~(\ref{m2ftsrlc}) and (\ref{mftsrlc}), the leading FTS features now become horizontal as clearly seen and the leading FSS regimes now have slopes of $-\gamma/\nu$ and $-\gamma/2\nu$ instead of $2\beta/\nu$ and $\beta/\nu$, respectively, in the FTS form in agreement with Table~\ref{charac}.

\begin{figure}
  \centering
  \includegraphics[width=\columnwidth,clip=true]{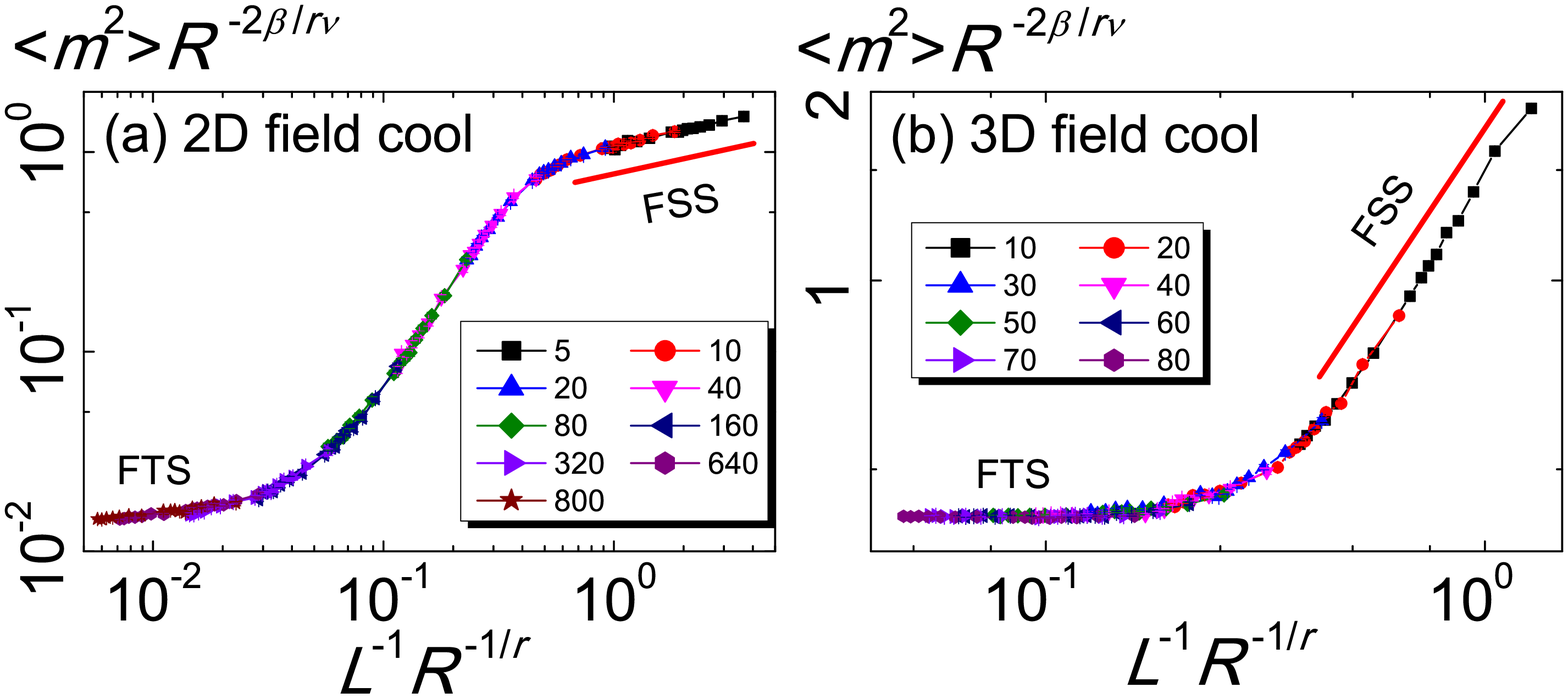}
  \vskip -2cm
  \caption{(Color online) \label{downh} FTS collapses of the squared magnetization at $T_c$ for cooling under an applied external field of $H=cR^{\beta\delta/r\nu}$, or $HR^{-\beta\delta/r\nu}=c$, a constant, which is chosen to be $0.1$ in 2D and $1$ in 3D. The legends give the lattice sizes used. The leading FTS sections now become horizontal, similar to the heating case in Figs.~\ref{m2R^}(a) and (c), rather than oblique as in Figs.~\ref{m2R^}(b) and (d), while the slopes of the leading FSS regimes remain as $2\beta/\nu$ given in Table~\ref{numc} and illustrated by the red line segments, identical for both heating and cooling. The extant $z$ values in both 2D and 3D have been used.}
\end{figure}
\begin{figure}
  \centering
  \includegraphics[width=\columnwidth,clip=true]{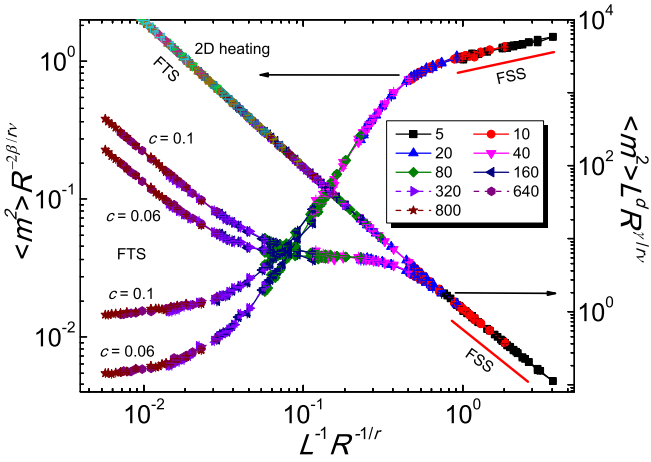}
%  \vskip -2cm
  \caption{(Color online) \label{2dhLR} FTS collapses of same data as in Fig.~\ref{downh}(a) for cooling under an applied external field in 2D. An additional data set of $c=0.06$ is also shown for comparison. Both sets use the extant $z$ value. Again for comparison, the data of Fig.~\ref{m2R^}(a) for heating, which subject to no external field, are also plotted with the right axis. The legend gives the lattice sizes used for the four curves with applied external fields and do not apply to the curve for heating. The two red line segments show the slopes of the leading FSS regime. The upper one is the same as the one in Fig.~\ref{downh} and has a slope $2\beta/\nu$, while the lower one has a slope $-\gamma/\nu$, both given in Table~\ref{numc}, according to the theory. The leading FTS regime in the right axis has a slope $-d$, which, in 2D, is close to $-\gamma/\nu$ in value. This is why the curve for heating appears nearly straight. The two horizontal arrows indicate the ordinate axis for the curves.}
\end{figure}
To confirm the origin of the extra feature in cooling, we plot in Fig.~\ref{downh} the squared magnetization at $T_c$ in cooling in the presence of a small applied external field. As $HR^{-\beta\delta/r\nu}$ is fixed to a constant $c$, the data of different rates and lattice sizes again collapse onto a single curve according to Eq.~(\ref{mftsh}) as can be seen in Fig.~\ref{downh}. In 3D, the rescaled curve for cooling in Fig.~\ref{downh}(b) now resembles well that for heating shown in Fig.~\ref{m2R^}(c). However, in 2D, from Figs.~\ref{downh}(a) and \ref{m2R^}(b), it appears that only an additional FTS regime emerges to the left, beyond which the remaining curve still resembles the cooling case in the absence of the applied field. To see this further, we present in Fig.~\ref{2dhLR} the collapses in $\langle m^2\rangle L^{d}R^{-\gamma/r\nu}$ versus $L^{-1}R^{-1/r}$. One sees that, although it appears somewhat oblique in the left axis, the leading FTS regime now has a slope near $-d$, which can be convinced by the 2D heating curve in the absence of the external field, in agreement with the characteristics listed in Table~\ref{charac}. In addition to this FTS section, there exist a horizontal section and an FSS section, both looking like the cooling case in the absence of the applied field as shown in Fig.~\ref{downm2R^}(a). Moreover, the horizontal section widens as $c$ decreases towards no field, in accompanying with the reducing $\langle m^2\rangle$. Therefore, we see that the intermediate section is indeed the finite-time-finite-size scaling regime in which the LPS scaling shows.

Several other conclusions can also be drawn from Fig.~\ref{2dhLR}. First, for sufficiently large $c$, the intermediate finite-time-finite-size scaling regime disappears and the cooling now behaves similar to heating as Fig.~\ref{downh}(b) for 3D shows. Second, the heating curve confirms the characteristics listed in Table~\ref{charac} when presented in $\langle m^2\rangle L^{d}R^{-\gamma/r\nu}$ versus $L^{-1}R^{-1/r}$. No intermediate finite-time-finite-size scaling regime appears. Third, the scaling function $F_{H0}$ is regular for small $c$ and thus can be expanded in it. This leads to the seemingly overlap of curves of different $c$ in the FSS regime because the first order in $c$ appears negligible in the figure. Fourth, the seemingly overlap of this FSS regime with the heating curve implies that in this regime, the scaling functions of cooling and heating are identical. Fifth, in the FTS regime, however, the scaling functions of heating and cooling are different. Indeed, as $c$ is lowered, the field-cooled curves tend to the zero-field-cooled curves in Figs.~\ref{m2R^}(b) and \ref{downm2R^}(a). Sixth, the separation of the curves of different $c$ in the FTS regime results of course from the field $H$. And it is the effect of the field that gives rise to the standard FTS regime. However, the detailed mechanism from which the intermediate regime arises is not yet clear.

\subsection{\label{eof}Effects on FTS}
As shown in the previous section, the scaling behavior in the FTS regime can be different in heating and cooling. This then calls for attention in employing FTS to determine critical exponents.

\begin{figure}
  \centering
  \includegraphics[width=\columnwidth,clip=true]{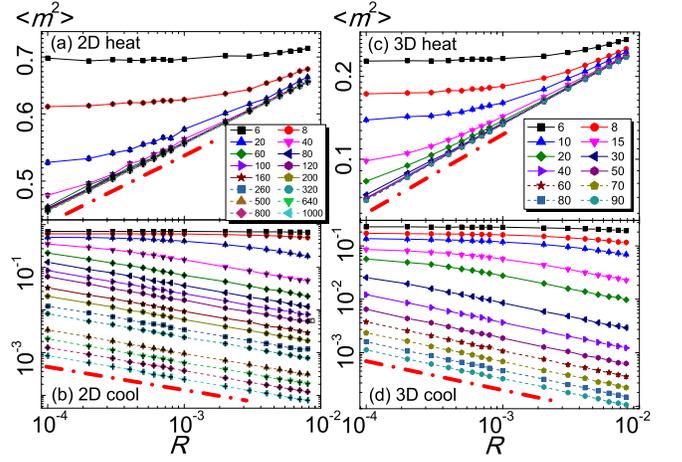}
  \caption{(Color online) \label{m2R} $\langle m^2\rangle$ versus $R$ for various lattice sizes given in the legends, which are identical with the ones in Fig.~\ref{m2R^}. The thick dash-dotted line segments have slopes of $2\beta/r\nu$ for heating and $-\gamma/r\nu$ for cooling from Eqs.~(\ref{m2ftsrlh}) and (\ref{m2ftsrlc}), respectively.}
\end{figure}

\begin{figure}
  \centering
  \includegraphics[width=\columnwidth,clip=true]{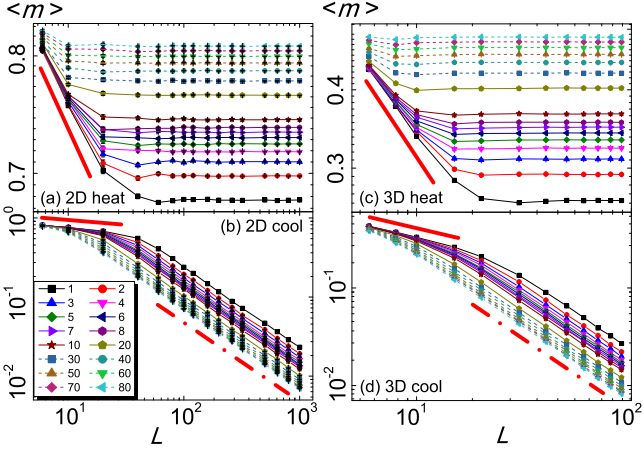}
  \caption{(Color online) \label{mL} $\langle m\rangle$ versus $L$ for various rates $R$. The legend is identical with the one in Fig.~\ref{m2L^} and gives the rate $R$ except in the form $10^4*R$ here. All the thick line segments have a slope $-\beta/\nu$ from Eqs.~(\ref{mftsrlh}) and (\ref{mftsrlc}), while both thick dash-dotted line segments have a slope of $-d/2$ from Eq.~(\ref{mftsrlc}).}
\end{figure}

To show this, we plot in Figs.~\ref{m2R} and \ref{mL} directly $\langle m^2\rangle$ versus $R$ and $\langle m\rangle$ versus $L$, respectively, without any rescaling contrast to Figs.~\ref{m2L^} to \ref{mR^} and Figs.~\ref{downm2R^} and \ref{downmR^}. From Fig.~\ref{m2R}(a) and (c), one sees that, for heating, the squared magnetization hardly depends on the lattice sizes for small $L$ and $R$ but converges to a straight line of a slope $2\beta/r\nu$, the usual critical exponent for $\langle m^2\rangle$ in the FTS regime for large sizes, in agreement with Eq.~(\ref{m2ftsrlh}). Conversely, one can then estimate this exponent from the slope. However, from Fig.~\ref{m2R}(b) and (d), for cooling, although $\langle m^2\rangle$ is again almost independent on $L$ for small $L$ and $R$, but it becomes parallel straight lines of a slope $-\gamma/r\nu$, instead of its representative exponent, according to Eq.~(\ref{m2ftsrlc}). Consequently, in cooling, if one would employ the dependence of $\langle m^2\rangle$ on $R$ to estimate its associated critical exponent, one would arrive at $-\gamma/r\nu$ instead of $2\beta/r\nu$! Ignoring the difference between heating and cooling would thus result in wrong results. Similar behaviors show in $\langle m\rangle$ as well.

To the contrast, from Fig.~\ref{mL}, one sees that the small $L$ and $R$ data converge to the leading FSS regime with a slope $-\beta/\nu$ irrespective of heating or cooling, although the FTS regimes are different as expected. The heating data are independent of $R$ for large $L$, whereas the cooling ones exhibit a slope $-d/2$ giving rise to the LPS scaling in the FTS regime in agreement with Eqs.~(\ref{mftsrlh}) and (\ref{mftsrlc}).

\subsection{\label{zvalues}The dynamic critical exponent $z$}
\begin{table*}
\centering
\caption {\label{tablez} The dynamic critical exponents estimated.}
\begin{ruledtabular}
\begin{tabular}{cccccccccccc}
 \multirow{2}{*}{dimension }& \multirow{2}{*}{condition}   & \multicolumn{3}{c}{FSS form}& \multicolumn{5}{c}{FTS form} & \multirow{2}{*}{average} & \multirow{2}{*}{average}\\\cline{3-5}\cline{6-10}
 & & $\langle m\rangle$ & $\langle m^2\rangle$ & average  & $\langle m\rangle$ & $\langle m^2\rangle$& $\langle m\rangle L^{d/2}$  & $\langle m^2\rangle L^{d}$ & average & & \\\hline
\multirow{2}{*}{2D}&heating& 2.123 & 2.124 &  2.1235(5) & 2.121 & 2.121 &- &- &2.121(1) &2.1223(9) &\multirow{2}{*}{2.155(3)}  \\%\cline{2-8}
  &cooling & 2.186 & 2.186 & 2.186(1)   & 2.194 & 2.188 & 2.189 &2.185 & 2.189(2) &2.188(2) &\\%\cline{1-12}
 \multirow{2}{*}{3D} & heating  &2.208 &2.211&2.210(2) & 2.210 &2.214 & -&- &2.212(2)&2.211(2) & \multirow{2}{*}{2.133(7)}\\%\cline{2-8}
  &cooling & 2.062 & 2.041 & 2.052(11) &2.066 &2.042 &2.064 &2.055 &2.057(6)&2.055(5) &\\
\end{tabular}
\end{ruledtabular}
\end{table*}
Now comes the dynamic critical exponent $z$. As pointed out in Sec.~\ref{chara}, the two-side method yields $z$ besides the division of the two regimes. The results are summarized in Table~\ref{tablez}. Each entry of $z$ so obtained is given to the third decimal place as the interval of its variation in the fitting is $0.001$. As has been pointed out above, the results from $\chi$ fluctuate a lot and we do not list them, possibly because of the deficiency of the numbers of the samples used. We have also listed the averages of all results from the FSS form (column 5) and FTS form (column 10) and both (column 11) in Table~\ref{tablez}. In Figs.~\ref{m2L^} to \ref{mR^} and Figs.~\ref{downm2R^} and \ref{downmR^}, we have used the corresponding $z$ values to draw them, while their averages, columns 5 and 10 in Table~\ref{tablez} have been employed to draw Figs.~\ref{chiL^} and \ref{chiR^}, respectively. We see from Table~\ref{tablez} that the results are quite consistent in that they are independent of both whether the FSS or the FTS forms and which variables are used to estimate them. This is also reflected in the high precisions of the averaged values in column 11. However, it can also be seen from Table~\ref{tablez} that the values of $z$ estimated from heating is different from those from cooling. Opposite trends are observed in 2D and 3D. For the former, the heating data are smaller than the cooling ones, while, for the latter, the heating data are larger than the cooling ones. If we accept the extant values of $z$, we see that the heating data are smaller than the extant one while the cooling data are larger in 2D, whereas both the heating and cooling $z$ are larger than the extant value in 3D. As $z$ is different in heating and cooling, the averaged values between heating and cooling listed in the last column in Table~\ref{tablez} can only serve as a reference.

To check whether these results are specific to the two-side method, we have also utilized the one-side method. We found however that this only affects the third decimal place of each $z$ value. We have also varied the range of $R$ in the one-side method. In this case, the $z$ values obtained vary a lot, from about $2.0$ to $2.25$ in both 2D and 3D, but without any appreciable systematic trends. This variation is understandable as a small fluctuation in one datum may change the fitting result. However, upon unbiased averages, averages that delete only large deviations beyond three standard deviations, the resultant $z$ values again agree with their corresponding ones listed in Table~\ref{tablez}, with the averaged values themselves differing only about 1 in the second decimal place in 2D and in the first decimal place for heating and less than 2 in the second decimal place for cooling in 3D. Nevertheless, the trends with respect to the extant values remain. Note that our results for cooling are close to the extant ones. Thus, we believe that the results are robust.

\begin{figure}
  \includegraphics[width=\columnwidth,clip=true]{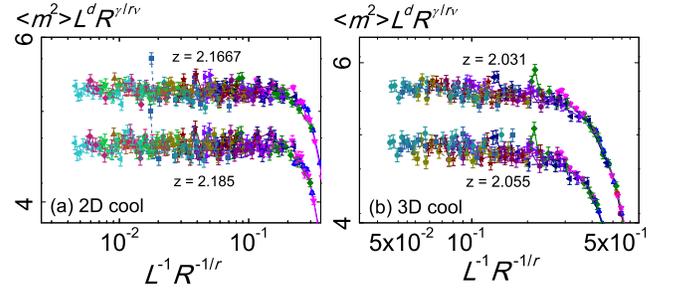}
  \vskip -2cm
  \caption{(Color online) \label{3ddownRz} Finite-time-finite-size scaling collapses of the same data as in Fig.~\ref{m2R^}(b) and (d) using the extant $z$ values instead of the present results. The original data in Fig.~\ref{m2R^}(b) and (d) are also present for comparison. For clarity of illustration, the data using the present $z$ values have been shifted downwards by $0.8$ in 2D and $0.9$ in 3D; otherwise, they cannot be distinguished from the other due to the small differences in $z$.}
\end{figure}
\begin{figure}
  \includegraphics[width=\columnwidth,clip=true]{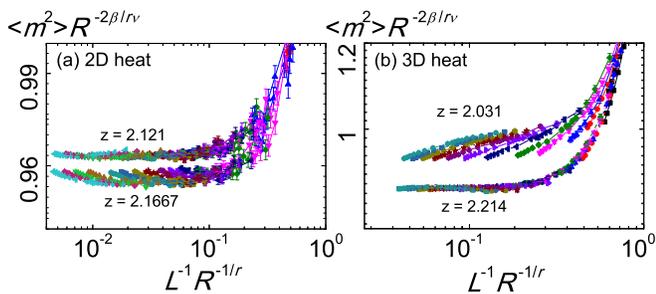}
  \vskip -2cm
  \caption{(Color online) \label{3dupRz} FTS collapses of the same data as in Fig.~\ref{m2R^}(a) and (c) using the extant $z$ values instead of the present results. The original data in Fig.~\ref{m2R^}(a) and (c) are also present for comparison. Clear size dependence is seen for the extant values.}
\end{figure}
To investigate the difference of our estimated $z$ values from the extant ones, we now compare the quality of the data collapses. We plot in Figs.~\ref{3ddownRz} and \ref{3dupRz} the same data collapses for $\langle m^2\rangle$ in Fig.~\ref{m2R^}(b) and (d) for cooling and (a) and (c) for heating, respectively, using the extant values of $z$ in 2D and 3D. From Fig.~\ref{3ddownRz} for cooling, one sees that the differences between the curves are only slight if any, in consistence with the small difference between the present values of $z$ and the extant ones. It may be seen that, in 2D, from the data of the large sizes, which appear on the left side of Fig.~\ref{3ddownRz}(a), the present results appear slightly better, whereas, in 3D, the extant results seem slightly better. Although larger sample sizes may be needed to confirm these differences, Fig.~\ref{3ddownRz} already gives support to our results and their quality in cooling. For heating, on the other hand, from Fig.~\ref{3dupRz}, it is clear that the present estimated values give rise to much better data collapses, especially in 3D. Therefore, if the quality of data collapses is the sole deciding factor, a point which seems to be supported by the results shown in Fig.~\ref{3ddownRz} for cooling to the extent of the present statistics, the present values of $z$ describe much better the numerical results in heating. But this then results in the different values of $z$ in heating and cooling.

A possible reason for the difference between the present results and the extant ones may arise from the correction to scaling~\cite{Wegner} which has not been taking into account. From Fig.~\ref{3dupRz}, it can be clearly seen that the data of different $R$ and $L$ do not collapse onto a single curve for the extant $z$ values. This indicates that the scaling functions are still $L$ and/or $R$ dependent if the extant $z$ values are true in heating. Although Fig.~\ref{chiR^}(c) and possibly Fig.~\ref{chiL^}(c) show that the present 3D $z$ in heating may also contain correction to scaling, the collapses in $\langle m^2\rangle$ and $\langle m\rangle$ appear quite good with little correction. On the other hand, Fig.~\ref{3ddownRz} shows that correction to scaling may be negligible in cooling both for the present $z$ values and the extant ones to the extent of the present statistics. So, it is not clear whether the present results underestimate the correction to scaling or there is indeed a difference in $z$ between heating and cooling as observed in the scaling behavior. We leave such questions for future studies.

\section{\label{sum}Conclusions}
We have studied in detail the relations between FTS and the KZ mechanism, LPS' scaling, and FSS. In particular, we have found that FTS and its essence, the controllable finite time scale, provides a detailed improved understanding of the KZ Mechanism. We have also shown that FTS not only accounts well for, but also improves, the scaling proposed by LPS, which incorporates the KZ mechanism with FSS. Further, we have found that the LPS universal scaling is applicable not to heating but only to to some observables in cooling under no or a small applied external field. In the latter case, the LPS scaling appears in an intermediate regime between the standard FTS and the FSS regimes.

We have demonstrated clearly the leading FTS regime and its crossover to the leading FSS regime for both heating and cooling and in both the FTS and the FSS forms. Detailed characteristics for this crossover behavior have been presented for several variables and confirmed unambiguously by numerical results for both heating and cooling. The dynamic critical exponent $z$ has accordingly been estimated for the 2D and 3D Ising models with the usual Metropolis dynamics. It gives rise to better data collapses than the extant values do at least in heating but takes on different values in heating and cooling in both 2D and 3D. Corrections to scaling are a possible origin for these differences.

In comparison with FSS, FTS exhibits subtlety and richer behaviors such as the LPS finite-time-finite-size scaling and its crossovers to FSS and standard FTS even for a single critical point of the Ising type. Consequently, caution should be excised in application of FTS to estimate the critical exponents.

\begin{acknowledgments}
We would like to thank Anders Sandvik for stimulating discussions.
This work was supported by NNSFC (10625420).
\end{acknowledgments}

%\newpage
%==========================  BIBLIOGRAPHY ============================%

%--------------------------------------------------%

\end{document}